\documentclass[11pt,english]{article}
\usepackage[T1]{fontenc}
\usepackage[latin9]{inputenc}
\usepackage{geometry}
\usepackage{amssymb}
\usepackage{babel}
\usepackage{jheppub}
\usepackage{graphicx}
\usepackage{color}

\begin{document}

\title{Reformulating the TBA equations for the quark anti-quark potential and their two loop expansion}

\author[a]{Zolt\'an Bajnok,}
\author[a]{J\'anos Balog,}
\author[b]{Diego H. Correa,}
\author[a]{\'Arp\'ad Heged\H us}
\author[b]{Fidel I. Schaposnik Massolo,}
\author[a]{and G\'abor Zsolt T\'oth}

\affiliation[a]{MTA Lend\"ulet Holographic QFT Group,
Wigner Research Centre,\\
H-1525 Budapest 114, P.O.B. 49, Hungary}

\affiliation[b]{Instituto de F\'{\i}sica La Plata, CONICET
 \\ Universidad Nacional de La Plata
 \\ C.C. 67, 1900 La Plata, Argentina}

\emailAdd{bajnok.zoltan@wigner.mta.hu}
\emailAdd{balog.janos@wigner.mta.hu}
\emailAdd{correa@fisica.unlp.edu.ar}
\emailAdd{hegedus.arpad@wigner.mta.hu}
\emailAdd{fidel.s@fisica.unlp.edu.ar}
\emailAdd{toth.gabor.zsolt@wigner.mta.hu}

\abstract{
The boundary thermodynamic Bethe Ansatz (BTBA) equations introduced
in \cite{Correa:2012hh,Dru-int-WL} to describe the cusp anomalous dimension
contain imaginary chemical potentials and singular boundary fugacities,
which make its systematic expansion problematic.
We propose an alternative formulation based on real chemical potentials and
additional source terms. We expand our equations to
double wrapping order and find complete agreement with the
direct two-loop gauge theory computation of the cusp anomalous dimension.}

\maketitle

\section{Introduction}

The cusp anomalous dimension is a very important physical quantity  in any gauge theory as it is related to various observables, such as the infrared divergences of massive scattering amplitudes, the energy emitted by an accelerated quark or even the quark anti-quark potential if the gauge theory is conformal.

In ${\cal N}=4$ super Yang-Mills, a boundary thermodynamic Bethe Ansatz (BTBA) has been derived for the exact computation
of the cusp anomalous dimension in the planar limit \cite{Dru-int-WL,Correa:2012hh}, which is a function of two cusp angles $\phi$ and $\theta$
and the 't Hooft coupling constant $g$. The proposed BTBA is similar
to the usual AdS/CFT thermodynamic Bethe Ansatz system for
closed strings
\cite{Arutyunov:2009zu,Gromov:2009tv,Bombardelli:2009ns,Gromov:2009bc,Arutyunov:2009ur,Arutyunov:2009ux}, with twisted boundary conditions \cite{vanTongeren:2013gva,Ahn:2011xq},
but includes an additional driving term originating from a boundary
dressing factor.

There are two particular features of this BTBA system that make the systematic expansion of the TBA equations quite subtle.
The first one is that the twist factors, which enter the TBA equations as chemical potentials, are imaginary for real cusp angles
$\phi$ and $\theta$. One problem of having imaginary chemical potentials is that Y-functions, although real, are not necessarily positive.
This seems to contradict the physical meaning of the ground state Y-functions in the Bethe Ansatz as ratios between densities of holes and
densities of particles and indicates that imaginary chemical potentials  might
correspond to \lq \lq excited states''. The other important feature is that the boundaries can emit and absorb particles with mirror kinematics.
These singular boundary fugacities give rise to integrals of logarithms with double poles in their arguments.
When computing those integrals square roots appear and one has to be careful to extract their signs.

In order to deal with the issues raised in the previous paragraph we present an alternative formulation of the BTBA.
In the first place,  we find such a domain of  parameters where the BTBA corresponds to a ground state, {\it i.e.} we consider all the chemical potentials to be real. This guarantees that the asymptotic Y-functions
are all positive and that the aforementioned square roots
can be safely taken with the positive sign. Since we are
eventually interested in the expectation value of physical Wilson loops with real cusp angles we will have to analytically continue
the chemical potentials to imaginary values in the final result. In so doing  singularities cross the integration contour, which  has to
be carefully investigated \cite{Bajnok:2007ep}. To avoid this, and concerning the singular boundary fugacities,
we will shift the contours of integration in such a way that all contributions sensitive to square root sign ambiguities
can be isolated. When shifting the contour of integrations upwards in the complex plane one crosses zero singularities of those logarithms
developing double poles. As a consequence of the shifts additional source terms are
generated while the remaining integrals with the shifted contours have no poles. The resulting BTBA  is of  an  excited type.

At this point it is important to emphasize that, although it seems more appropriate to work with real chemical potentials
and eventually analytically continue from that, it is still possible to work with imaginary chemical potentials,
provided the signs of the additional source terms with origin in the singular fugacities are chosen properly.
Following the physical intuition, in \cite{Correa:2012hh} the signs in the integral giving the cusp anomalous dimension were chosen such that
in the limit of $\phi\to\pi$ all contributions are negative. It is not difficult to see that if one adopts the same sign choice
for all the integrals in the TBA equations with singular fugacities the final answer for the 2-loop cusp anomalous dimension is the same as
the analytical continuation of the answer with real chemical potentials. We expect this to be true to any loop order.

In the original formulation of the cusp anomalous dimension BTBA, the prescription of the sign choice is useful only
for the analytical computation of the TBA integrals because the sign choice affects only the pole contribution
and not the full integrals. Now, by shifting the contour of integration, we will isolate the pole contribution
from the integral and the prescription of assigning precise signs while working with real angles is more
easily implemented. As a consequence, the reformulated BTBA equations appear to be appropriate for a numerical study
of the anomalous dimensions for real cusp angles, as done for the Konishi operator \cite{Gromov:2009zb}.

After considering this reformulation of the BTBA for the cusped Wilson loop we will study its asymptotic expansion and solve the integral equations to second order, {\it i.e.} we will compute double wrapping
corrections. This will allow us to extract the 2-loop cusp anomalous dimension from the BTBA system.

Let us recall what  the gauge theory observable under study is. We will consider a locally supersymmetric Wilson loop,
which includes a coupling with the scalar fields of the theory through a unitary vector $\vec n$:
\begin{equation}
W \sim {\rm tr} \left[P e^{i\oint A\cdot dx + \oint \vec\Phi\cdot\vec n |dx|}\right]\,.
\end{equation}
We consider the contour to be a line with a cusp angle $\phi$ and take $\vec n$ and $\vec n'$
to define the couplings with the scalar fields before and after the cusp.

\begin{figure}[h]
\centering
\def\svgwidth{10cm}
 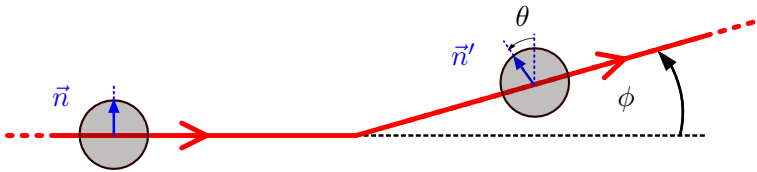
\caption{Generalized cusped Wilson line}
\label{cuspedline}
\end{figure}

The expectation value of such a Wilson loop develops logarithmic divergences coming from the cusp \cite{Polyakov:1980ca,Korchemsky:1991zp}
\begin{equation}
\langle W \rangle \sim e^{-\Gamma(\phi,\theta,g)\log\frac{\epsilon_{\rm IR}}{\epsilon_{\rm UV}}}\,,
\end{equation}
with $\epsilon_{\rm IR}$ and $\epsilon_{\rm UV}$ infrared and ultraviolet cutoffs, respectively.

In the planar limit, the cusp anomalous dimension, $\Gamma(\theta,\phi,g)$ can be expanded in powers of the 't Hooft coupling as follows
\begin{equation}
\Gamma(\theta,\phi,g)=\sum_{k=0}^{\infty}\Gamma_{k}(\theta,\phi)g^{2k}\:,
\label{Gammacusp}
\end{equation}
where the $\theta$ dependence of each loop order is of the form
\begin{equation}
\Gamma_{k}(\theta,\phi)=\sum_{n=1}^{k}\left(\frac{\cos\phi-\cos\theta}{\sin\phi}\right)^{n}
\gamma_{k}^{(n)}(\phi)\:.
\label{Gammak}
\end{equation}
The one loop term in  this weak coupling expansion is simply
\begin{equation}
\gamma_{1}^{(1)}=\frac{\phi}{2}\:.
\label{gamma1}
\end{equation}
There are two terms contributing at two loops\footnote{Explicit results for $\Gamma_3$ and $\Gamma_4$ are also known \cite{Correa:2012nk,Henn:2013wfa}.}. The simpler one is
\begin{equation}
\gamma_{2}^{(1)}=\frac{\phi}{12}(\phi^{2}-\pi^{2})\:,
\end{equation}
while the more complicated is
\begin{equation}
\gamma_{2}^{(2)}=\frac{1}{4}\left[\zeta_{3}-\mbox{Li}_{3}(e^{2i\phi})
+i\phi\left(\mbox{Li}_{2}(e^{2i\phi})+\frac{\pi^{2}}{6}\right)-i\frac{\phi^{3}}{3}\right]\:.
\end{equation}
The complicated term can be characterized as
\begin{equation}
\gamma_{2}^{(2)}(0)=\gamma_{2}^{(2)\prime}(0)=0\:,\quad\qquad
\gamma_{2}^{(2)\prime\prime}(\phi)=\frac{\phi}{2}\cot\phi\:.
\end{equation}
In the following we recover this stunningly simple expression from the weak coupling expansion of the BTBA equations.

\section{BTBA equations}

In this section we recall the canonical BTBA equations and their asymptotic
solution. We suggest a way to deal with integrands having a double pole at the origin by shifting the integration contours. Finally,  we rewrite the TBA equations into the hybrid form, which makes the formulation of the weak coupling expansion easier.

\subsection{Canonical equations}

Our starting point is the set of canonical TBA equations \cite{Correa:2012hh,Dru-int-WL}
describing the cusp anomalous dimension of the generalized Maldacena-Wilson
loops with
the insertion of a local operator at the cusp,
\begin{equation}
{\cal O} = Pe^{\int_{\cal C}(i A_\mu \dot{x}^\mu + |\dot{x}|\vec\Phi \cdot \vec n)dt} Z^L
           e^{i\int_{\cal C'}(i A_\mu \dot{x'}^\mu + |\dot{x'}|\vec\Phi \cdot \vec n')dt}\,,
\label{WLwithinsertion}
\end{equation}
where $\dot{x}^\mu \dot{x}'_\mu = \cos\phi$ and $\vec n\cdot\vec n'=\cos\theta$.

For the purposes of this paper
we have rewritten the equations in the conventions used previously for
the mirror TBA description of states with periodic boundary
conditions \cite{AFS}. The new feature of TBA equations for Maldacena-Wilson
loops is the presence of driving terms originating from the boundary dressing
phase and driving terms proportional to external chemical potentials.
Mirror TBA equations with chemical potentials have been discussed in
\cite{CFT} and they also appear in the context of beta and gamma
deformed models \cite{Ahn:2011xq,deLeeuw:2012hp}.

The unknowns (Y-functions) are associated to nodes of the left-right
symmetric $AdS_5\times S^5$ Y-system: $Y_Q, Q=1,2,\dots$ for the massive
nodes, $Y_\pm$ for the fermionic modes, $Y_{m|v}$ and $Y_{m|w}$ ($m=1,2,\dots$)
for the two different type of magnonic nodes\footnote{In order to shorten
the notation we abbreviated the magnonic Y-function from $Y_{m|vw}$ to
$Y_{m|v}$.}. We have one TBA equation
for every node:
\begin{equation}
\begin{aligned}
\ln Y_Q&=-2\psi Q-R\tilde\varepsilon_Q+\ln M_Q+\sum_{Q^\prime=1}^\infty L_{Q^\prime}
\star K_{{\rm sl}(2)}^{Q^\prime Q}
+2\sum_{m=1}^\infty {\cal L}_m\star K_{vwx}^{mQ}\,,    \\
&\qquad\qquad +2{\cal L}_- \,\hat\star\, K_-^{yQ}
+2{\cal L}_+ \,\hat\star\, K_+^{yQ}\,,\\
\ln Y_\pm&=f-t -\sum_{Q=1}^\infty L_Q\star K^{Qy}_\pm+\sum_{m=1}^\infty\left(
{\cal L}_m-\tilde{\cal L}_m\right)\star K_m\,,\\
\ln Y_{m|v}&=2mf -\sum_{Q=1}^\infty L_Q\star K^{Qm}_{xv}+\sum_{m^\prime=1}^\infty
{\cal L}_{m^\prime}\star K_{m^\prime m}+\left({\cal L}_--{\cal L}_+\right)\,
\hat\star\, K_m\,,\\
\ln Y_{m|w}&=2mt+\sum_{m^\prime=1}^\infty
\tilde{\cal L}_{m^\prime}\star K_{m^\prime m}+\left({\cal L}_--{\cal L}_+\right)\,
\hat\star\, K_m\,.
\end{aligned}
\label{cano}
\end{equation}
Here we have used the notations
\begin{equation}
\begin{split}
L_Q&=\ln(1+Y_Q),\qquad
{\cal L}_m=\ln\left(1+\frac{1}{Y_{m|v}}\right),\qquad
\tilde{\cal L}_m=\ln\left(1+\frac{1}{Y_{m|w}}\right),\\
{\cal L}_\pm&=\ln\left(1-\frac{1}{Y_\pm}\right),\qquad\quad
\tilde\varepsilon_Q=\ln\frac{x^{[-Q]}}{x^{[Q]}}\,,
\end{split}
\end{equation}
and the definition of the various kernels can be found in ref. \cite{AFS}.
The parameter $R$ is defined as $R=2(L+1)$, where $L$ is the number of
local scalar operators inserted at the cusp and $M_Q$ is coming from the
boundary dressing phase:
\begin{equation}
M_Q=\exp\left\{i\chi\left(x^{[-Q]}\right)+i\chi\left(1/x^{[Q]}\right)
-i\chi\left(1/x^{[-Q]}\right)-i\chi\left(x^{[Q]}\right)\right\}.
\end{equation}
The analytic function $\chi(z)$ is defined through the integral \cite{Correa:2012hh,Dru-int-WL}
\begin{equation}
\Phi(z)=\oint\limits_{|\omega|=1}\frac{{\rm d}\omega}{2\pi}\frac{1}{\omega-z}
\ln\frac{\sinh\pi g\left(\omega+1/\omega\right)}
{\pi g\left(\omega+1/\omega\right)},\qquad |z|\not=1\,,
\end{equation}
as
\begin{equation}
\begin{split}
\chi(z)&=\Phi(z)\qquad\quad|z|>1,\\
\chi(z)&=\Phi(z)-i\ln\frac{\sinh\pi g\left(z+1/z\right)}
{\pi g\left(z+1/z\right)}
\qquad\quad|z|<1.
\end{split}
\end{equation}
For later purposes, using the identity
$
\Phi(z)=\Phi(0)-\Phi(1/z)
$
we write $M_Q$ in the alternative form
\begin{equation}
M_Q(u)=\exp\left\{2i\Phi\left(x^{[-Q]}\right)
+2i\Phi\left(1/x^{[Q]}\right)-2i\Phi\left(0\right)\right\}
\,\frac{\pi^2(g^2u^2+Q^2)}{\sinh^2\pi gu}.
\end{equation}
In this form it is clearly seen that all $M_Q(u)$ have a double pole at $u=0$.

Finally, the TBA equations (\ref{cano}) contain driving terms proportional
to the chemical potentials $\psi$, $f$ and $t$. These are identified with
the geometrical and internal angles as follows \cite{Correa:2012hh,Dru-int-WL}:
\begin{equation}
\psi=i(\pi-\phi),\qquad f=i(\phi-\pi),\qquad t=i(\theta-\pi),
\end{equation}
{\it i.e.} all chemical potentials are imaginary. This is similar to the cases
studied in \cite{Ahn:2011xq,deLeeuw:2012hp}. In this paper we will perform an analytic
continuation in these parameters and for the moment we treat all three
parameters as independent. Note that this three-parameter set of
driving terms is the most general one \cite{CFT} we can add to the TBA
equations without changing the corresponding Y-system relations.

After having found the solution of the TBA integral equations (\ref{cano}),
the energy of the ground state is given by
\begin{equation}
E_0(L) =-\frac{1}{4\pi}\sum_{Q=1}^\infty\int_{-\infty}^\infty{\rm d}u
\,\frac{{\rm d}\tilde P_Q}{{\rm d}u}\,L_Q(u),
\label{Gamma}
\end{equation}
where
\begin{equation}
\tilde P_Q=gx^{[-Q]}-gx^{[Q]}+iQ.
\end{equation}
The energy of the ground state $E_{0}(L)$ is the anomalous dimension of the operator (\ref{WLwithinsertion}).
This energy will be expanded as follows:
\begin{eqnarray}
E_{0}(L) &=& E_{0}^{(0)}(L) + E_{0}^{(2)}(L)+ \cdots
\nonumber
\\
&=& E_{0}^{(0,2L+2)}(L) g^{2L+2}+E_{0}^{(0,2L+4)}(L) g^{2L+4}+ \cdots + E_{0}^{(2,4L+4)}(L) g^{4L+4}+ \cdots
\end{eqnarray}
where $a$ in $E_{0}^{(a)}(L)$ is associated to the wrapping order.
The cusp anomalous dimension (\ref{Gammacusp}) is then obtained by setting $L=0$, {\it i.e.} $\Gamma = E_0(0)$.

\subsection{Asymptotic solution, master formula and the leading term}

The asymptotic solution valid for large volume ($R\to\infty$) or weak coupling ($g\to0$) can be obtained by calculating the (super)trace of the
double row transfer matrix.
This solution must satisfy the TBA equations in the asymptotic limit, where the massive nodes are small
and the terms containing $L_Q$ can be neglected. In this limit all Y-functions (except the massive ones $Y^o_Q$) are constants and the TBA equations
simplify drastically. For the TBA equations (\ref{cano}) with real chemical potentials $f$, $t$ and $\psi$ one finds (see Appendix \ref{asymsol} for details)
\begin{equation}
\begin{split}
Y^o_\pm & =\frac{\cosh f}{\cosh t},\qquad\quad
Y^o_{m|w}=\frac{\sinh mt\,\sinh(m+2)t}{\sinh^2 t},\qquad
Y^o_{m|v}=\frac{\sinh mf\,\sinh(m+2)f}{\sinh^2f},\\
Y^o_Q & =4e^{-2(f+\psi)Q}\frac{\sinh^2 Q f}{\sinh^2 f}\,(\cosh f-\cosh t)^2
\left(\frac{x^{[Q]}}{x^{[-Q]}}\right)^{2L+2}\,M_Q.
\end{split}
\label{ASY}
\end{equation}
We used an upper index $^o$ to indicate that they are the asymptotic values. Clearly,
for real chemical potentials the Y-functions are all positive as expected.
In the  TBA language ground state Y-functions are given by the formula
\begin{equation}
Y=\frac{\rm density\ of\ holes}{\rm density\ of\ particles},
\end{equation}
which is a manifestly positive quantity. On the contrary, for real angles, {\it i.e.} for imaginary chemical potential,  the \(Y_{m\\ |v    }\) and  \(Y_{m\\ |w    }\) functions would not be everywhere positive, thus they should correspond to some excited state TBA.

~

The formula for the cusp anomalous dimension is a sum of
integrals of the following generic form:
\begin{equation}
I=\int_{-\infty}^\infty{\rm d}u\,{\cal K}(u)\ln Z(u)\,,
\label{I}
\end{equation}
with integrands having a double pole in the argument of the logarithm:
\begin{equation}
Z(u)=1+\frac{\Lambda(u)}{u^2}\,.
\end{equation}
For (\ref{Gamma}) we need this integral with
\begin{equation}
{\cal K}(u)=-\frac{1}{4\pi}\,\frac{{\rm d}\tilde P_Q}{{\rm d}u},\qquad\qquad
\Lambda(u)=u^2Y_Q(u)\,.
\end{equation}
All ${\cal K}(u)$ and $\Lambda(u)$ are even, real analytic functions, moreover
$\Lambda(u)$ is asymptotically small of order O$(\epsilon^2)$. We will use
the small parameter $\epsilon$ to characterize the smallness of terms in the
asymptotic limit (for $R\to\infty$ or $g\to0$).

Naively the integral $I$ is O$(\epsilon^2)$ but because of the presence
of the double pole $\ln Z$ is not uniformly O$(\epsilon^2)$ and $I$
turns out to be only O$(\epsilon)$. A similar situation has been encountered
previously in the boundary Sinh-Gordon model \cite{Bajnok:2006dn,Bajnok:2007ep}. The treatment of
integrals of the form (\ref{I}) can be borrowed from that calculation, see also  Appendix A.

Although (\ref{I}) is convergent, to avoid problems coming from the
fact that the integration contour goes through the double pole, we
shift the integration contour by $i\eta$. The new integration contour is
parallel to the real axis, away from it by the finite amount $\eta$.
Because of the smallness and evenness of $\Lambda$, there is a zero of
$Z(u)$ at $u=iu_0$ on the imaginary axis close to the origin:
\begin{equation}
Z(iu_0)=0,\qquad\qquad u_0={\rm O}(\epsilon).
\label{Ziuo}
\end{equation}
We have to take into account the contribution of this zero when performing
the shift of the contour:
\begin{equation}
I=-2\pi iS(iu_0)+\int_{-\infty
+i\eta}^{\infty+i\eta}{\rm d}u\,{\cal K}(u)
\ln Z(u),
\label{Ishift}
\end{equation}
where $S$ is the odd primitive of ${\cal K}$:
\begin{equation}
S^\prime(u)={\cal K}(u),\qquad\qquad S(0)=0.
\end{equation}
Let us emphasize that equation (\ref{Ishift}) together with the quantization condition (\ref{Ziuo}) is completely equivalent to
(\ref{I}). It is, however, much more suited for expansion in $\epsilon$.

Away from the double pole we can safely expand $\ln Z$ and because of the
smallness of $u_0$ also the term containing $S$. The result is
\begin{equation}
I=2\pi {\cal K}(0)\sqrt{\Lambda(0)}
+\int_{-\infty+i\eta}^{\infty+i\eta}{\rm d}u\,{\cal K}(u)
\frac{\Lambda(u)}{u^2}+{\rm O}(\epsilon^3),
\label{master}
\end{equation}
where we have used
\begin{equation}
u_0=\sqrt{\Lambda(0)}+{\rm O}(\epsilon^3).
\end{equation}

We will now use the master formula (\ref{master}) to calculate the leading
order contribution to the cusp anomalous dimension in the weak coupling
expansion. This comes from the first term in (\ref{master}). Noting that
\begin{equation}
\tilde P^\prime_Q(0)=g+{\rm O}(g^3),\qquad
\Lambda_Q(0)=\left(\frac{g^2}{Q^2}\right)^{R-1}C_Q^2
\left(1+{\rm O}(g^2)\right)\,,
\end{equation}
where
\begin{equation}
C_Q=\frac{\cosh f-\cosh t}{\sinh f}\,\left\{{\rm e}^{-\psi Q}-
{\rm e}^{-(2f+\psi)Q}\right\}\,,
\end{equation}
we find
\begin{equation}
E_0(L)=\left(g^2\right)^{L+1}E^{(0,2L+2)}_0(L)+{\rm O}\left((g^2)^{L+2}\right)\,,
\end{equation}
with leading order coefficient
\begin{equation}
E^{(0,2L+2)}_0(L)=-\frac{1}{2}\,\sum_{Q=1}^\infty\,\frac{|C_Q|}{Q^{2L+1}}\,.
\label{problematic}
\end{equation}

There is a problem with equation (\ref{problematic}). Taking the absolute value of $C_Q(\phi,\theta)$
for all values of the cusp angles cannot be correct, because it would not lead to the 1-loop
cusp anomalous dimension (\ref{Gammak}),(\ref{gamma1}) when setting $L=0$. Equation (\ref{problematic}) can only be valid in
a safe domain, {\it i.e.} for certain values of the chemical potential.
For all the other values outside the safe domain, the correct answer can be obtained by analytic continuation.
The safe domain is parametrized by three independent real positive chemical potentials
$\psi$, $f$ and $t$ satisfying
\begin{equation}
\psi>0,\qquad\quad f>t>0,
\end{equation}
which moreover leads to a solution of the BTBA equation such that all the Y-functions are positive.

Accepting this prescription, we can now do the calculation of $E^{(0)}_0(L)$ in the safe domain.
To obtain $\Gamma_1$ we set $L=0$,
\begin{equation}
\Gamma_1=-\frac{1}{2}\,\frac{\cosh f-\cosh t}{\sinh f}\,\left\{
{\cal F}(\psi)-{\cal F}(2f+\psi)\right\}\,,
\end{equation}
where
\begin{equation}
{\cal F}(\psi)=\sum_{Q=1}^\infty\frac{{\rm e}^{-Q\psi}}{Q}=
-\ln\left(1-{\rm e}^{-\psi}\right)\,.
\end{equation}
This function has a cut along the negative real axis so the analytic continuation from positive real to nonzero imaginary poses no problem
and we get
\begin{equation}
\Gamma_1= \frac{\cos\phi-\cos\theta}{2\sin\phi} \phi\,,
\label{Gamma1}
\end{equation}
when making the analytic continuation
\begin{equation}
\psi\to i(\pi-\phi),\qquad\quad f\to i(\phi-\pi),\qquad\quad
t\to i(\theta-\pi)\,.
\label{ancon}
\end{equation}

In \cite{Correa:2012hh}, a different prescription was used to obtain exactly the same result for $\Gamma_1$ as in (\ref{Gamma1}):
in equation (\ref{problematic}) the absolute value bars were simply removed with an additional sign of $(-1)^Q$ to ensure
that in the strict limit $\phi\to\pi$ all $Q's$ contribute negatively. Both prescriptions will successfully reproduce
the 2-loop cusp anomalous dimension when going beyond the leading asymptotic order. As we will discuss below, the prescription
employed in \cite{Correa:2012hh} in combination with the shift of the integration contour, would be useful to numerically study the
TBA equations for real cusp angles.

\subsection{Hybrid equations}

We will now map the canonical TBA equations (\ref{cano}) to an equivalent
set of equations, the hybrid equations \cite{AFS}. Since the mathematical
transformation affects only the convolution and chemical potential terms
its derivation is identical to the one presented in \cite{vanTongeren:2013gva} and will
not be repeated here. Our hybrid equations are as follows:
\begin{align}
\ln Y_{m|w}&=\tilde R_{m+1}\star s+\tilde R_{m-1}\star s+\delta_{m1}\ln
\left(\frac{1-1/Y_-}{1-1/Y_+}\right)\,\hat\star\, s,\\
\ln Y_{m|v}&=-L_{m+1}\star s +R_{m+1}\star s+R_{m-1}\star s+\delta_{m1}\ln
\left(\frac{Y_--1}{Y_+-1}\right)\,\hat\star\, s,\\
\ln\frac{Y_+}{Y_-}&=\sum_{Q=1}^\infty L_Q\star K_{Qy},\\
\ln Y_+Y_-&=-\sum_{Q=1}^\infty L_Q\star K_Q+2\sum_{Q=1}^\infty L_Q\star
K_{xv}^{Q1}\star s+2R_1\star s-2\tilde R_1\star s,\\
\ln Y_Q&=-2(f+\psi)Q-R\tilde\varepsilon_Q+\ln M_Q+\sum_{Q^\prime=1}^\infty L_{Q^\prime}\star K_s^{Q^\prime Q}\nonumber\\
&+2R_1\star s\,\hat\star\, K_{yQ}+2R_{Q-1}\star s+\ln
\left(1-\frac{1}{Y_+}\right)\left(1-\frac{1}{Y_-}\right)\,\hat\star\, K_{yQ}\\
&+\ln\left(\frac{1-1/Y_-}{1-1/Y_+}\right)\,\hat\star\, K_Q -2\ln\left(\frac
{Y_--1}{Y_+-1}\right)\,\hat\star\, s\star K_{vwx}^{1Q}.\nonumber
\end{align}
Here we introduced the notations
\begin{equation}
R_m=\ln(1+Y_{m|v}),\qquad \tilde R_m=\ln(1+Y_{m|w}),\qquad R_0=\tilde R_0=0\,,
\end{equation}
and
\begin{equation}
s(u)=\frac{g}{4\cosh\frac{\pi g u}{2}}\,,
\end{equation}
for the universal TBA kernel. We also made the abbreviation
\(
K_s^{Q^\prime Q}= K_{{\rm sl}(2)}^{Q^\prime Q}+
2s\star K_{vwx}^{Q^\prime-1 Q} \). For the definition of the other kernel
functions we refer to \cite{AFS}.
In this hybrid form of the equations
only the sum of the two chemical potentials, $f+\Psi$, is present
explicitly. The other parameters appear in the large $m$ asymptotics
of the magnonic Y-functions:
\begin{equation}
\ln Y_{m|w}=2mt+O(1),\qquad  \qquad\ln Y_{m|v}=2mf+O(1).
\label{eq:mus}
\end{equation}

\section{Reformulating the  BTBA equations}

The aim of this section is to use the master formula (\ref{Ishift})
to reformulate the hybrid BTBA equations into a form which allows a systematic large volume, expansion and also numerical studies.
 We denote the appearing pole contributions as
\begin{equation}
D_\alpha^\beta (iu_Q)=-2 \pi i S_\alpha^\beta(iu_Q)\,,
\end{equation}
where $\alpha$ and $\beta $   refer to the various kernels we convolve with $L_Q$. The obtained equations read as follows:
\begin{align}
\ln Y_{m|w}&=\tilde R_{m+1}\star s+\tilde R_{m-1}\star s+\delta_{m1}\ln
\left(\frac{1-1/Y_-}{1-1/Y_+}\right)\,\hat\star\, s,\\
\ln Y_{m|v}&=-D_{s}(iu_{m+1})-L_{m+1}\star_{
\eta }s+R_{m+1}\star s+R_{m-1}\star s+\delta_{m1}\ln
\left(\frac{Y_--1}{Y_+-1}\right)\,\hat\star\, s,\\
\ln\frac{Y_+}{Y_-}&=\sum_{Q=1}^\infty (D_{Qy}(iu_{Q})+L_Q\star_{\eta} K_{Qy}),\\
\ln Y_+Y_-&=-\sum_{Q=1}^\infty (D_{Q}(iu_{Q})
+L_Q\star_{\eta} K_Q)+2\sum_{Q=1}^\infty (D^{Q1}_{xvs}(iu_{Q})+L_Q\star_{\eta}
K_{xv}^{Q1}\star s)\nonumber \\ &\qquad +2R_1\star s-2\tilde R_1\star s,\\
\ln Y_Q&=-2(f+\psi)Q-R\tilde\varepsilon_Q+\ln M_Q
+\sum_{Q^\prime=1}^\infty \Bigl(D_{s}^{Q'Q}(iu_{Q'})+
L_{Q^\prime}\star_{\eta}K_s^{Q^\prime Q}\Bigr)\nonumber\\
&\quad +2R_1\star s\,\hat\star\, K_{yQ}+2R_{Q-1}\star s+\ln
\left(1-\frac{1}{Y_+}\right)\left(1-\frac{1}{Y_-}\right)\,\hat\star\,
K_{yQ}\label{ReBTBA}\\
&\quad +\ln\left(\frac{1-1/Y_-}{1-1/Y_+}\right)\,\hat\star\, K_Q -2\ln\left(\frac
{Y_--1}{Y_+-1}\right)\,\hat\star\, s\star K_{vwx}^{1Q}.\nonumber
\end{align}
where we denoted the shifted convolution by
\begin{equation}
f\star_{\eta}K=\int_{-\infty+i\eta}^{\infty+i\eta}du\, f(u)K(u,v)\:.
\end{equation}
The location of the source terms, $u_{Q}$, are determined from the
equation
\begin{equation}
1+Y_{Q}(iu_{Q})=0\:.\label{eq:uQ}
\end{equation}
The energy with the shifted contour takes the form
\begin{equation}
E_{0}(L)=\frac{i}{2}\tilde{P}_{Q}(iu_{Q})-\sum_{Q=1}^{\infty}
\int_{-\infty+i\eta}^{\infty+i\eta}\frac{\rm{d}u}{4\pi}\frac{{\rm d}
\tilde{P}_{Q}}{\rm{d}u}L_{Q}\:.\label{eq:energy}
\end{equation}
The whole system is similar to the system of excited state TBA equations.

In the following we perform an asymptotic large volume/weak coupling
expansion of these reformulated hybrid BTBA equations.

\subsection{Asymptotic expansion}

We expand the reformulated hybrid TBA equations to leading and next-to-leading order
in the parameter $\epsilon=e^{-\epsilon_{Q}R}$. The expansions of the $Y$-functions
are denoted as
\begin{equation}
Y=Y^o(1+y+\dots)\:.
\end{equation}
We solve iteratively the BTBA equations together with the quantization
condition (\ref{eq:uQ}) for $u_{Q}=u_{Q}^{(0)}+u_{Q}^{(1)}+\dots$.

At leading order the massive nodes are exponentially small, so neglecting
them splits the $Y$-system into two independent subsystems which have constant asymptotic solutions.
These constant values determine the LO exponentially small expressions for the massive nodes $Y^o_{Q}$
which determine $u_{Q}^{(0)}.$
At LO the solutions $Y^o$ are the ones presented in section 2.
The constant $Y^o_{n|v},\, Y^o_{m|w}$ functions are the same as
one of the wings of the
deformed $O(4)$ model \cite{Ahn:2011xq} and can be written as
\begin{equation}
Y^o_{m\vert v}=[m]_f[m+2]_{f}\,;\quad Y^o_{n|w}=[n]_{t}[n+2]_{t}\,,
\end{equation}
where
\begin{equation}
[n]_{c}=q^{n-1}+q^{n-3}\dots+q^{3-n}+q^{1-n}=\frac{q^{n}-q^{-n}}{q-q^{-1}}=
\frac{\sinh nc}{\sinh c}\,,\qquad\qquad q=e^{c}\,.
\end{equation}

Comparing these results to the $\gamma$-deformed theories, \cite{Ahn:2011xq},
we can observe that the $Y^o_{n|w}$ functions are basically
the same, while the $Y^o_{n|v}$ functions got deformed, too.
The resulting equations look as if we had analyzed a system in deformed AdS space, such that
the TBA equations for the other $su(2)$ part were also twisted,
similarly to \cite{deLeeuw:2012hp,vanTongeren:2013gva}.

The previous $Y^o_{n|v},\, Y^o_{m|w}$   asymptotic  solutions, altogether with the fermionic $Y^o_{\pm}$
\begin{equation}
Y^o_{\pm}=\sqrt{\frac{1+Y^o_{1|v}}{1 +Y^o_{1|w}}}=\frac{[2]_{f}}{[2]_{t}}
=\frac{\cosh f}{\cosh t}\:.
\end{equation}
led to the following asymptotic solution for the massive Y-function
\begin{equation}
Y^o_{Q}=[Q]_{f}^2([2]_{f}-[2]_{t})^{2}M_{Q}e^{-2(f+\Psi)Q-R
\tilde{\epsilon}_{Q}}\:.
\end{equation}

Now plugging back $Y^o_{Q}$ into eq. (\ref{eq:uQ}) we obtain
the asymptotic location of $u_{Q}=u_{Q}^{(0)}+\dots$:
\begin{equation}
u_{Q}^{(0)}=m_{Q}[Q]_{f}([2]_{f}-[2]_{t})e^{-(f+\Psi)Q-
\frac{R}{2}\tilde{\epsilon}_{Q}(0)}>0\:,\qquad\qquad m_{Q}=
\sqrt{\lim_{u\to0}u^{2}M_{Q}(u)}\:.
\end{equation}
At leading order the integral term can be neglected in the energy
formula (\ref{eq:energy}) and the full LO correction is
\begin{equation}
E_{0}^{(0)}(L)=-\frac{1}{2}\sum_{Q=1}^{\infty}\frac{{\rm d}
\tilde{P}_{Q}}{\rm{d}u}
(0)u_{Q}^{(0)}=
-\frac{1}{2}([2]_{f}-[2]_{t})\sum_{Q=1}^{\infty}
\frac{{\rm d}
\tilde{P}_{Q}}{\rm{d}u}(0)[Q]_{f}m_{Q}
e^{-(f+\Psi)Q-\frac{R}{2}\tilde{\epsilon}_{Q}(0)}\,\,.\label{eq:LO}
\end{equation}

\subsection{NLO correction}

At NLO we have to include the integral term in the energy formula
(\ref{eq:energy}) and additionally we have to calculate the NLO correction
of $u_{Q}$.

Here we focus on the calculation of the NLO correction to
$u_{Q}=u_{Q}^{(0)}+u_{Q}^{(1)}+\dots$
.We use the equation
\begin{equation}
1+Y^o_{Q}(iu_{Q})(1+y_{Q}(iu_{Q}))=0\:,\label{eq:uQNLO}
\end{equation}
where $y_{Q}$ should be determined from the linearized TBA equations:
\begin{eqnarray}
y_{Q} & = & 2\pi u_{Q'}K_{s}^{Q'Q}+2A_{1\vert v}y_{1|v}\star s\,
\hat{\star}K_{yQ}+2A_{Q-1\vert v}y_{Q-1|v}\star s\nonumber \\
 &  & \,\,\,\,\,\,\,\,\,\,\,\,\,\,\,\,\,\,\,\,\,\,\,-2\frac{y_{-}
 -y_{+}}{1-\frac{1}{Y^o_{+}}}\hat{\star}s\star K_{vx}^{1Q}
 +\frac{y_{-}-y_{+}}{(Y^o_{+}-1)}\hat{\star}K_{Q}
 +\frac{y_{-}+y_{+}}{(Y^o_{+}-1)}\hat{\star}K_{yQ}\:,\\
y_{+}+y_{-} & = & 2\left(A_{1\vert v}y_{1|v}-A_{1\vert w}y_{1|w}\right)
\star s-4\pi u_{Q}K_{xv}^{Q1}\star s-2\pi u_{Q}K_{Q}\:,\label{linTBAy++y-}\\
y_{+}-y_{-} & = & 2\pi u_{Q}K_{Qy}\:,\label{linTBAy+-y-}\\
y_{m|v} & = & \left(A_{m-1\vert v}y_{m-1|v}+A_{m+1\vert v}y_{m+1|v}\right)
\star s-2\pi u_{m+1}\star s+\delta_{m1}\frac{y_{-}-y_{+}}
{1-\frac{1}{Y^o_{+}}}\hat{\star}s\:,~~~~\\
y_{n|w} & = & \left(A_{n-1\vert w}y_{n-1|w}+A_{n+1\vert w}y_{n+1|w}\right)
\star s+\delta_{n1}\frac{y_{+}-y_{-}}{1-Y^o_{+}}\hat{\star}s\,.
\end{eqnarray}
where $A_{m\vert v}=\frac{Y^o_{m\vert v}}{1+Y^o_{m\vert v}}=
\frac{[m]_{f}[m+2]_{f}}{[m+1]_{f}^{2}}$
and $A_{n\vert w}=\frac{Y^o_{n\vert w}}{1+Y^o_{n\vert w}}=
\frac{[n]_{t}[n+2]_{t}}{[n+1]_{t}^{2}}$.
Here any combination of the form $u_{Q}K_{.}^{Q.}$
is understood as $u_{Q}K_{.}^{Q.}(0,v)$.

The solution of this system of linearized equations can be written
into the form
\begin{equation}
y_{Q}=u_{Q'}\mathcal{M}_{Q'Q}\:.
\end{equation}
The correction to $u_{Q}$ can be calculated from eq. (\ref{eq:uQNLO})
to NLO as
\begin{equation}
-u_{Q}^{2}Y^o_{Q}(iu_{Q})(1+y_{Q}(iu_{Q}))=u_{Q}^{2}\,\,.
\end{equation}
As the lhs is an even function of $u_{Q}$ we can keep the LO term
only
\begin{equation}
\lim_{u\to0}(u^{2}Y^o_{Q}(u))(1+y_{Q}(iu_{Q}^{0}))=
(u_{Q}^{(0)})^{2}+2u_{Q}^{(0)}u_{Q}^{(1)}+O(\epsilon^{3})\,\,.
\end{equation}
Taking into account the LO solution leads to
\begin{equation}
u_{Q}^{(1)}=\frac{1}{2}y_{Q}(iu_{Q}^{(0)})u_{Q}^{(0)}=
\frac{1}{2}u_{Q'}^{(0)}\mathcal{M}_{Q'Q}u_{Q}^{(0)}\:.
\end{equation}
The calculation of $\mathcal{M}_{Q'Q}$ is a generalization of that was
performed for the double wrapping correction in \cite{Ahn:2011xq}
for the $\gamma$-deformed theories. There it was shown that the $\mathcal{M}_{QQ'}$
quantity can   be calculated in two alternative ways: either from the TBA equations or directly from the scattering   and   twist matrix.
Both calculations are presented in Appendix C and result in the
same expression
\begin{eqnarray}
y_{Q_{2}} & = & u_{Q_{1}}\Bigg\{2\pi K_{sl(2)}^{Q_{1}Q_{2}}
+\frac{2[2]_{t}}{[2]_{f}-[2]_{t}}\frac{1}{i}\partial_{u_{1}}
\ln a_{1}^{Q_{1}Q_{2}}(u_{1},u_{2})+\frac{2}{[Q_{1}]_{f}[Q_{2}]_{f}}
\frac{1}{i}{\cal K}_{f}^{Q_{1}Q_{2}}\\
 &  & +\frac{2}{[2]_{f}-[2]_{t}}\frac{1}{i}\partial_{u_{1}}
 \left[\frac{[Q_{2}-1]_{f}}{[Q_{2}]_{f}}\ln a_{2}^{Q_{1}Q_{2}}(u_{1},u_{2})
 +\frac{[Q_{1}-1]_{f}}{[Q_{1}]_{f}}\ln a_{2}^{Q_{2}Q_{1}}(u_{2},u_{1})^{\star}
 \right]\Bigg\}\nonumber
\end{eqnarray}
where
\begin{equation}
a_{1}^{Q_{1}Q_{2}}(u_{1},u_{2})=A^{-1}\quad;\qquad a_{2}^{Q_{1}Q_{2}}(u_{1},u_{2})
=AB\quad;\qquad a_{2}^{Q_{2}Q_{1}}(u_{2},u_{1})^{\star}=AB^{-1}\label{eq:a1a2}
\end{equation}
with
\begin{equation}
A=\frac{x_{1}^{-}-x_{2}^{+}}{x_{1}^{+}-x_{2}^{-}}\sqrt{\frac{x_{1}^{+}}{x_{1}^{-}}}
\sqrt{\frac{x_{2}^{-}}{x_{2}^{+}}}\quad;\qquad B=
\frac{1-x_{1}^{+}x_{2}^{+}}{1-x_{1}^{-}x_{2}^{-}}
\sqrt{\frac{x_{1}^{-}}{x_{1}^{+}}}\sqrt{\frac{x_{2}^{-}}{x_{2}^{+}}}\:;
\quad x_{i}^{\pm}=x^{[\pm Q_{i}]}\label{eq:AB}
\end{equation}
 Furthermore
\begin{equation}
\mathcal{K}_{f}^{Q_{1}Q_{2}}=\sum_{j=1}^{Q_{1}-1}[j]_{f}[Q_{21}+j]_{f}
K_{Q_{21}+2j}\qquad;\qquad K_{Q}=\partial_{u_{1}}\ln\frac{u_{1}-u_{2}-i
\frac{Q}{g}}{u_{1}-u_{2}+i\frac{Q}{g}}\quad.\label{eq:Kf}
\end{equation}
where we assumed that $Q_{21}=Q_{2}-Q_{1}\geq0$. The matrix $\mathcal{M}_{Q_{1}Q_{2}}$
is symmetric. Alternatively
\begin{equation}
\mathcal{K}_{f}^{Q_{1}Q_{2}}=\sum_{j=0}^{Q_{1}-2}[Q_{21}+2j+1]_{f}
\sum_{k=1}^{Q_{1}-j-1}K_{Q_{21}+2j+2k}\:.
\end{equation}
Combining this result with the integral term we obtain the full NLO
correction:
\begin{equation}
E_{0}^{(2)}(L)=-\frac{1}{4}\sum_{Q,Q'}\frac{{\rm d}
\tilde{P}_{Q}}{\rm{d}u} (0)
\mathcal{M}_{QQ'}u_{Q}^{(0)}u_{Q'}^{(0)}-\sum_{Q=1}^{\infty}
\int_{-\infty+i\eta}^{\infty+i\eta}\frac{du}{4\pi}\frac{{\rm d}
\tilde{P}_{Q}}{\rm{d}u}Y^o_{Q}\:.
\end{equation}
In the next section we perform a weak coupling expansion of this result,
together with the LO correction (\ref{eq:LO}), in order to get the 2-loop
cusp anomalous dimension.

\section{Weak coupling expansion}

In the following we perform the weak coupling expansion of the LO
and NLO corrections for real chemical potentials and continue the result
back to the physical angles. First we analyze at which orders of $g^{2}$
the various terms contribute. The detailed expansion of the various functions can be found in  Appendix D, here we summarize the result.

In order to expand the LO term
\begin{equation}
E_{0}^{(0)}(L)=-\frac{1}{2}\sum_{Q=1}^{\infty}\frac{\mathrm{d}\tilde{P}_{Q}}{\mathrm{d}u}(0)u_{Q}^{(0)}\quad;
\qquad u_{Q}^{(0)}=([2]_{f}-[2]_{t})[Q]_{f}m_{Q}e^{-(f+\Psi)Q-\frac{R}{2}\tilde{\epsilon}_{Q}(0)}\,\,.\label{eq:LO}
\end{equation}
we need
\begin{equation}
\frac{\mathrm{d}\tilde{P}_{Q}}{\mathrm{d}u}(0)=g -\frac{2g^3}{Q^2}+\dots \quad;\quad e^{-\tilde{\epsilon}_{Q}(0)}=\frac{g^{2}}{Q^{2}}-\frac{2g^{4}}{Q^{4}}
+\dots\quad;\quad m_{Q}=\frac{Q}{g}+\frac{\pi^{2}gQ}{3}+\dots
\end{equation}
where ellipses denotes higher order terms in $g^2$. As a consequence the
expansion of $u_{Q}^{(0)}$ is
\begin{equation}
u_{Q}^{(0)}=([2]_{f}-[2]_{t})[Q]_{f}e^{-(f+\Psi)Q}\left(\frac{g}{Q}\right)^{2L+1}
\left(1+g^{2}\left(\frac{\pi^{2}}{3}-\frac{2(L+2)}{Q^{2}}\right)+O(g^{4})\right)
\label{eq:uQweak}
\end{equation}
and the leading order correction of $E_{0}(L)$ starts at $g^{2L+2}$:
\begin{equation}
E_{0}^{(0)}(L)=E_{0}^{(0,2L+2)}(L)g^{2L+2}+E_{0}^{(0,2L+4)}(L)g^{2L+4}+\dots
\end{equation}
This correction is the only one until the NLO correction
\begin{equation}
E_{0}^{(2)}(L)=-\frac{1}{4}\sum_{Q,Q'}\frac{\mathrm{d}\tilde{P}_{Q}}{\mathrm{d}u}(0)
\mathcal{M}_{QQ'}u_{Q}^{(0)}u_{Q'}^{(0)}-\sum_{Q=1}^{\infty}\int_{-\infty+i\eta}^{\infty+i\eta}
\frac{du}{4\pi}\frac{\mathrm{d}\tilde{P}_{Q}}{\mathrm{d}u}Y_{Q}^{o}\:.
\label{eq:NLO}
\end{equation}
starts to contribute. The $g$-dependence of the first term can be
calculated from $\mathcal{M}_{QQ'}\propto g$ and using that
\begin{equation}
M_{Q}(u)=\frac{\pi^{2}(\tilde{P}^{2}+Q^{2})}{\sinh^{2}\pi\tilde{P}}+\dots\quad;
\qquad e^{-\tilde{\epsilon}_{Q}(\tilde{P})}=\frac{g^{2}}{\tilde{P}^{2}+Q^{2}}+\dots
\end{equation}
 we can see that the integral scales the same way. This means that
\begin{equation}
E_{0}^{(2)}(L)=E_{0}^{(2,4L+4)}(L)g^{4(L+1)}+\dots
\end{equation}
thus the large volume expansion of the TBA equations goes in the powers
of $e^{-(L+1)\tilde{\epsilon}_{Q}}$, {\it i.e.} a new term appears at the
order $g^{2n(L+1)}$.

In the following we concentrate on the cusp anomalous dimension ,
$E_0(0)=\Gamma,$
at order $g^{4}$ (as we already calculated the leading $g^{2}$ correction
in section 2). This amounts to calculating the $g^{2}$ correction
in (\ref{eq:uQweak}), $E_{0}^{(0,4)}=\Gamma_2^{(0)}$, and evaluating the leading
$g-$expansion of (\ref{eq:NLO}), $E_{0}^{(2,4)}=\Gamma_2^{(2)}$.

The contribution $\Gamma_2^{(0)}$
can be calculated as
\begin{eqnarray}
\Gamma_2^{(0)}
& = & -([2]_{f}-[2]_{t})\sum_{Q=1}^{\infty}\left(\frac{\pi^{2}}{6Q}-\frac{2}{Q^{3}}\right)
[Q]_{f}e^{-(f+\Psi)Q}\nonumber \\
 & = & -\frac{(\cosh f-\cosh t)}{\sinh f}\left[\frac{\pi^{2}}{6}\log\frac{1-e^{-f}}{1-e^{f}}-2(
 \mbox{Li}_{3}(e^{f})-\mbox{Li}_{3}(e^{-f}))\right]\label{eq:E(0,2)}\\
 & = & -\frac{(\cos\phi-\cos\theta)}{\sin\phi}\frac{\phi}{6}\left[\pi^{2}-2\phi^{2}\right]\:.\nonumber
\end{eqnarray}
where in the last line we substituted the physical angles. Observe
that scaling out $\sinh (f)$ from the sum  in the first line of (\ref{eq:E(0,2)})
leads to a sum, which
 vanishes for $f=0$ just as its
first derivatives and the second derivative is proportional to the
one loop result.

Every contribution coming from $E_{0}^{(0)}$ is proportional to
$\frac{(\cos\phi-\cos\theta)}{\sin\phi}$.
In particular, $\Gamma_2^{(0)}$
contributes to $\gamma_{2}^{(1)}.$
Let us denote this contribution by $\gamma_{2}^{(1a)}$ and by $\gamma_{2}^{(1b)}$
the contribution coming form $\Gamma_2^{(2)}$.
The $\theta$ angle dependence
of $\Gamma_2^{(2)}$
can be decomposed as
\begin{equation}
\Gamma_2^{(2)}
=
\frac{(\cos\phi-\cos\theta)}{\sin\phi}\gamma_{2}^{(1b)}
+\frac{(\cos\phi-\cos\theta)^{2}}{\sin^{2}\phi}\gamma_{2}^{(2)}
\end{equation}
The term $\gamma_{2}^{(1b)}$ comes from the $a_{1},a_{2},a_{2}^{\star}$
term of $\mathcal{M}$ and contributes as:
\begin{eqnarray}
&&  -2([2]_{f}-[2]_{t})\sum_{Q_{1},Q_{2}=1}^{\infty}\frac{[Q_{1}]_{f}}{Q_{1}}
\frac{[Q_{2}]_{f}}{Q_{2}}\left\{ [2]_{f}\frac{1}{i}\partial_{u_{1}}\log a_{1}+
\frac{[Q_{1}-1]_{f}}{[Q_{1}]_{f}}\frac{1}{i}\partial_{u_{1}}\log a_{2}^{\star}
\right\} = \nonumber \\
 &&  -2([2]_{f}-[2]_{t})\sum_{Q_{1},Q_{2}=1}^{\infty}\frac{[Q_{1}]_{f}}{Q_{1}}\frac{[Q_{2}]_{f}}{Q_{2}}
 \left\{ -\frac{[2]_{f}}{Q_{1}}+2\frac{[Q_{1}-1]_{f}}{Q_{1}[Q_{1}]_{f}}\right\} \end{eqnarray} from which it follows that
\begin{equation}
 \gamma_2^{(1b)} =  -2
 \phi\left(\frac{\phi^{2}}{2}-\frac{\pi^{2}}{6}\right)
\end{equation}
Combining the two terms $\gamma_2^{(1a)} $ and  $\gamma_2^{(1b)} $
we indeed arrive at $\gamma_{2}^{(1)}=\gamma_{2}^{(1a)}+\gamma_{2}^{(1b)}$,
which agrees with the gauge theory result.

The remaining $\gamma_2^{(2)}$
term can be further decomposed into
the integral part, $\gamma_{2}^{(2a)},$ and the term coming from $\mathcal{M}$:
$\gamma_{2}^{(2b)}$.
The integral term is
\begin{eqnarray}
& & -g^{-4}\sum_{Q=1}^{\infty}\int_{-\infty+i\eta}^{\infty+i\eta}\frac{du}{4\pi}
\frac{\mathrm{d}\tilde{P}_{Q}}{\mathrm{d}u}[Q]_{f}^{2}([2]_{f}-[2]_{t})^{2}M_{Q}e^{-2(f+\Psi)Q-2
\tilde{\epsilon}_{Q}}=\nonumber \\
  & & \sum_{Q=1}^{\infty}[Q]_{f}^{2}([2]_{f}-[2]_{t})^{2}e^{-2(f+\Psi)Q}
 \int_{-\infty+i\eta}^{\infty+i\eta}\frac{dq}{4\pi}\frac{\pi^{2}}{\sinh^{2}\pi q}\,\frac{1}{q^{2}+Q^{2}}
\end{eqnarray}
We perform the integral by residues
\begin{equation}
-\int_{-\infty+i\eta}^{\infty+i\eta}\frac{dq}{4\pi}\frac{\pi^{2}}{\sinh^{2}\pi q}\,\frac{1}{q^{2}+Q^{2}}
=\frac{1}{2Q}\Psi_{1}(Q)-\frac{1}{4Q^{3}}\;.
\end{equation}
The sum we encounter is
\begin{equation}
S(\phi)=\sum_{Q=1}^{\infty}\sinh^{2}(fQ)e^{-2(f+\Psi)Q}\left(\frac{1}{2Q}\Psi_{1}(Q)-
\frac{1}{4Q^{3}}\right)
\end{equation}
Actually it is easier to perform the sum for the derivative of $S(\phi)$:
\begin{equation}
S=0\qquad;\qquad S^{\prime}=-\frac{1}{4}\phi\left(\pi-\phi\right)\cot(\phi)
\end{equation}
Thus we arrive at
\begin{equation}
\gamma_{2}^{(2a)}=
\int_{0}^{\phi}\varphi
\left(\pi-\varphi\right)\cot(\varphi)d\varphi\:.
\end{equation}
The most complicated term is $\gamma_{2}^{(2b)}$.
This very technical
calculation
can be found in Appendix D.

It turns out that it is easier to calculate the derivatives of
$\gamma_{2}^{(2b)}(\phi)$
than the quantity itself.  One finds that
\begin{equation}
\gamma_{2}^{(2b)}(0)=0\quad ;\quad \gamma_{2}^{(2b)
\prime}(0)=-\pi \quad ; \quad \gamma_{2}^{(2b)
\prime \prime}= (\frac{5\phi}{2}-\pi)\cot(\phi)+\frac{\phi(\pi-\phi)}
{\sin(\phi)^2}
\end{equation}
Combining  the two terms
$\gamma_{2}^{(2)}=\gamma_{2}^{(2b)}+\gamma_{2}^{(2b)}$ we indeed recover
the two loop gauge theory result.

\section{Imaginary chemical potentials and numerical implementation of BTBA}
In this short section we comment on how the analytical continuation in the chemical potentials   (\ref{ancon}) can be done at the level of the  reformulated  BTBA equations (\ref{ReBTBA}).

In doing the analytical continuation in the angles no singularity will cross the integration contour as we already shifted it away from the \(\pm  u_Q \)s.
What it instead  changes is the solution of (\ref{eq:uQ}).  Depending on the continued angle $\phi$ some of the $u_Q$ should be taken
 on the upper (+), while some other  on the lower half plane (-). Concretely, on the asymptotic solution we have to take
\begin{equation}
u_{Q}^{(0)}=(-1)^Qm_{Q}[Q]_{i
\phi}([2]_{i\phi}-[2]_{i\theta})e^{-
\frac{R}{2}\tilde{\epsilon}_{Q}(0)}\qquad;\qquad m_{Q}=
\sqrt{\lim_{u\to0}u^{2}M_{Q}(u)}>0\:.
\label{imuQ}
\end{equation}
With this $(-1)^Q$ prescription we can expand the BTBA equations for
real angles and compare the result with the analytically continued
analogue  obtained  from real chemical potentials. We did this calculation at the two loop level and the results agreed.  This also explains the one loop calculation and the square root choice in \cite{Dru-int-WL,Correa:2012hh}.

Using this  $(-1)^Q$ prescription we can also solve the reformulated  BTBA equations (\ref{ReBTBA})
numerically.
We start the iterative solution for large volumes, $R$, with the  asymptotic solution of the Y functions (\ref{ASY})
and using the  asymptotic $u^{(0)}_Q$ as given  in (\ref{imuQ}).
We then follow numerically how the various functions and quantization  positions  evolve during the iteration.

\section{Conclusion}

In this paper we reformulated the BTBA equations which describe the
cusp anomalous dimension $\Gamma(\theta,\phi,g)$ in the ${\cal N}=4$
SYM theory. We obtained our equations by shifting the integration
contours and by explicitly including the crossed pole singularities as
extra source terms. Thus our BTBA equations are of the form of
excited state TBA equations.

We needed this reformulation at least for two reasons. On one hand,
real (physical) angles $\theta$ and $\phi$ lead
to imaginary chemical potentials, which result in non-positive $Y$
functions characteristic for excited states. On the other hand,
singular boundary fugacities make the expansion of the original BTBA
equations problematic.

We started to shift the contour from a domain when all $Y$ functions
were positive and we certainly described the ground state. We identified
such domain for real chemical potentials, {\it i.e.} for imaginary angles.

The continuation of the equations from imaginary to real angles
leads to the change
of the sign of some of the source terms depending on the angle $\phi$.
This method explains the sign choice in \cite{Dru-int-WL,Correa:2012hh} and the
resulting equations can be used for numerical studies.

The reformulated BTBA equations, due to the shifted contour, allow a
systematic large volume expansion and we think that  a similar method can be used for any
BTBA system with singular boundary fugacities. To test these ideas
we expanded our
equation at double wrapping order and compared the result to explicit
two loop gauge theory calculations.

Our result is a non-trivial precision test for double wrapping
corrections in the weak
coupling limit of AdS/CFT TBA systems.
Similar double wrapping corrections have been computed before in
\cite{deLeeuw:2012hp,Ahn:2011xq} for the $\gamma$ deformed theories.
However, for all those cases there is no explicit gauge
theory computation to compare to.
In the present case the double wrapping corrections contribute to the 2-loop cusp anomalous dimension
and we have found a complete agreement with the explicit perturbative results.

In  \cite{Correa:2012hh}
some double and triple wrapping term  were checked by
comparing to the exact result for the Bremsstrahlung function
\cite{Correa:2012at,Fiol:2012sg}, but only in the small cusp angle limit.
Remarkably, in this very particular limit, the BTBA was exactly solved in \cite{Gromov:2012eu}
and agreement with the exact Bremsstrahlung function was observed.
Although very impressive as a precision test, that result only depends on
 the residue of the pole of the
 reflection factor. Since all integrals were  dominated by double pole
contributions, that computation probed the boundary dressing factor to all orders
in the 't Hooft coupling but only in $u\to 0$ limit. In contrast, by reproducing the 2-loop
cusp anomalous dimension from solving the BTBA system to double wrapping order, we have probed
the boundary dressing factor in the weak coupling limit  for all values of $u$.

Gauge theory calculations are available also for the three, $\Gamma_3$,
 and four loop cusp anomalous dimensions, $\Gamma_4$
\cite{Correa:2012nk,Henn:2013wfa}. It would be particularly interesting
to recover their results (or even go beyond) from expanding the
BTBA equations further.
Probably to achieve this aim one has to adopt the
formulation based on the $P-\mu$ system \cite{Gromov:2013pga,
Gromov:2013qga}.

Another direction for future research is to recover the single
component BES integral equation for the cusp anomalous dimension
\cite{Beisert:2006ez}. We believe that our reformulated BTBA
equations with real chemical potentials are particularly useful
in this respect.

\section*{Acknowledgements}

We thank Nadav Drukker, Laszlo Palla and Amit Sever for useful discussions.
Z.B., J.B., A.H. and G.ZS.T.  were supported by a Lend\"ulet Grant.
Z.B. was supported by OTKA 81461, while
 A.H. by a Bolyai Scholarship and by OTKA 109312.
D.H.C. and F.I.S.M were supported by CONICET and
grants PICT 2010-0724 and PICT 2012-0417.
We thank IEU, Seoul for hospitality,
where part of this work was performed.
This investigation has also been supported in part by the
Hungarian National Science Fund OTKA (under K 77400)
and by the Hungarian-Argentinian bilateral grant
TÉT-10-1-2011-0071.

\appendix

\section{Regularizing BTBA's with singular fugacities}

In this Appendix we explain how one can regularize BTBA's with singular
fugacities. These singularities appear whenever in the strip geometry
both boundaries can emit/absorb virtual particles and make it difficult to develop
a systematic infra-red expansion of the ground-state BTBA equations.
Our primary example is the sinh-Gordon theory with Dirichlet boundary
conditions on both ends of the strip.

\subsection{Sinh-Gordon boundary TBA }

The sinh-Gordon theory is one of the simplest integrable models. It
contains one single particle with mass $m$ and scattering matrix
\begin{equation}
S=\frac{\sinh\theta-i\sin B\pi}{\sinh\theta+i\sin B\pi}=-(-B)_{\theta}
(1+B)_{\theta}\quad,\qquad(x)_{\theta}=\frac{\sinh(\frac{\theta}{2}
+\frac{i\pi x}{2})}{\sinh(\frac{\theta}{2}-\frac{i\pi x}{2})}\:.
\end{equation}
In the Lagrangian formulation a free boson is perturbed with the potential
$V(\varphi)=\frac{m^{2}}{b^{2}}(\cosh b\varphi-1)$ and $B=\frac{b^{2}}{8\pi+b^{2}}$.

We analyze the theory on the interval of size $L$ with Dirichlet
boundary conditions: $\varphi_{-}$ on the right and $\varphi_{+}$
on the left boundaries. These boundary conditions are integrable,
and represent how the particles reflect off from the boundary:
\begin{equation}
R_{\pm}(\theta)=\frac{\left(\frac{1}{2}\right)_{\theta}\left(1
-\frac{B}{2}\right)_{\theta}}{\left(\frac{3}{2}-\frac{B}{2}\right)_{\theta}}
\frac{\left(\pm\frac{iB\varphi_{\pm}}{b}-\frac{1}{2}\right)_{\theta}}{
\left(\pm\frac{iB\varphi_{\pm}}{b}+\frac{1}{2}\right)_{\theta}}\:.\label{eq:RDir}
\end{equation}
For $\varphi_{\pm}\neq0$ these reflection factors have poles at
$\theta=i\frac{\pi}{2}$:
\begin{equation}
R_{\pm}(\theta)=i\frac{g_{\pm}^{2}}{2\theta-i\pi}+\dots\qquad;
\qquad g_{\pm}=2\sqrt{\cos\frac{\pi B}{4}\cos\frac{\pi(1-B)}{4}}
\tan\Bigl(\frac{2\pi B}{b}\varphi_{\pm}\Bigr)\:.\label{eq:g}
\end{equation}
The quantities $g_{\pm}$ are the strengths of the virtual particle
absorbtions and emissions by the boundaries. We expect them to be
analytic functions of the boundary parameters. As only their square
appear in the reflection factors we have to be careful how to extract
their signs. We choose $g>0$ for $\varphi>0$ and analytically extend
it by (\ref{eq:g}) for $\varphi<0$. In the following we will be
interested in the ground state energy $E_{0}(L)$ on the strip.

\subsubsection{BTBA equations}

For $g=0$ a BTBA equation can be derived for the ground-state energy
\cite{LeClair:1995uf}:
\begin{equation}
E_{0}(L)=-m\int_{-\infty}^{\infty}\frac{d\theta}{4\pi}\cosh\theta\,
\ln\left(1+\lambda(\theta)\,\, e^{-\epsilon(\theta)}\right)\label{E0++}
\end{equation}
where $\lambda(\theta)=R_{+}(\frac{i\pi}{2}-\theta)\, R_{-}(\frac{i\pi}{2}+\theta)$.
The pseudo energy, $\epsilon(\theta)$, satisfies the BTBA equation
\begin{equation}
\epsilon(\theta)=2mL\cosh\theta-\int_{-\infty}^{\infty}
\frac{d\theta^{'}}{2\pi}\varphi(\theta-\theta^{'})\,\,
\ln\left(1+\lambda(\theta')\,\, e^{-\epsilon(\theta')}\right)\label{TBA-1}
\end{equation}
where $\varphi(\theta)$ is the logarithmic derivative of the bulk
scattering matrix: $\varphi(\theta)=\frac{1}{i}\frac{d}{d\theta}\ln S(\theta)$.

As was observed in \cite{Dorey:1997yg} the equation is also valid
for non-vanishing $g$'s, whenever $\varphi_{-}\varphi_{+}>0$. In
this case the ground state configuration is a  \lq \lq symmetric'' function,
contrary to the $\varphi_{-}\varphi_{+}<0$ case where it is \lq \lq anti-symmetric''.
To describe the anti-symmetric bound state one can continue analytically
in $\varphi_{-}$. In so doing two zeros of the logarithm
\begin{equation}
1+\lambda(\theta_{0})e^{-\epsilon(\theta_{0})}=0\label{eq:theta0}
\end{equation}
will cross the integration contour, which have to be added as additional
source terms, and we basically describe an excited state \cite{Bajnok:2007ep}.
Once we have the correct equations we can try a systematic large volume
expansion. However, as $\lambda$ has a double pole at the origin
the logarithm cannot be expanded and one has to be very careful even
in extracting the leading order correction \cite{Bajnok:2006dn,Bajnok:2007ep}.

To avoid these complications we develop a reformulation of the BTBA
equations, which allows a systematic large volume expansion. It amounts
to shifting the contours of integrations slightly above the real axis,
above $\theta_{0}$, and to picking up its contributions.

We start by assuming that $\varphi_{-}\varphi_{+}>0$ and integrate
the BTBA equation by parts:
\begin{equation}
\epsilon(\theta)=2mL\cosh\theta+\int_{-\infty}^{\infty}\frac{d
\theta^{'}}{2\pi i}[\ln(S(\theta-\theta')-\ln S(\theta)]\,\,
\frac{d}{d\theta'}\ln\left(1+\lambda(\theta')\,\, e^{-\epsilon(\theta')}\right)\:.
\end{equation}
 In order for the integral to be well-defined, we subtracted $\ln S(\theta)$
to ensure a finite integrand at $\theta'=0$. By shifting the contour
we pick up the residue term at $\theta_{0}$. To have a form similar
to the original equation we integrate by parts again:
\begin{equation}
\epsilon(\theta)=2mL\cosh(\theta)-\ln\Bigl(\frac{S(\theta)}{S(\theta-
\theta_{0})}\Bigr)-\int_{-\infty+i\eta}^{\infty+i\eta}\frac{d\theta^{'}}{2\pi}
\varphi(\theta-\theta^{'})\,\,\ln\left(1+\lambda(\theta')\,\,
e^{-\epsilon(\theta')}\right)\:,\label{eq:BTBAeta}
\end{equation}
where $\eta$ is arbitrary in the interval $\frac{\pi}{2}>\pi
B>\eta>\Im m(\theta_{0})$.
Doing the same manipulation in the energy term we obtain

\begin{equation}
E_{0}(L)=\frac{im}{2}\sinh\theta_{0}-m\int_{-\infty+i\eta}^{\infty
+i\eta}\frac{d\theta}{4\pi}\cosh(\theta)\,\ln\left(1+\lambda(\theta)\,\,
e^{-\epsilon(\theta)}\right)\:.\label{eq:E0eta}
\end{equation}
Equations (\ref{eq:BTBAeta}) and (\ref{eq:theta0}) determine $\theta_{0}$
and $\epsilon(\theta)$ simultaneously, which leads to the ground
state energy via (\ref{eq:E0eta}).

These equations are valid for
any $\varphi_{-}\varphi_{+}$ but we have to take care of the sign
of $\theta_{0}$ in solving (\ref{eq:theta0}). For $\varphi_{-}\varphi_{+}>0
$ we  choose the $\Im m(\theta_0)>0$ solution as follows from the contour shift, while for
$\varphi_{-}\varphi_{+}<0$ we have to take the $\Im m(\theta_0)<0$
one, which can be understood by following the  movement of $\theta_0 $ under analytical continuation, or
can be seen from the asymptotical solution what we calculate in the following.

\subsubsection{Large volume expansion}

We now develop a systematic large volume expansion. The idea is to
solve (\ref{eq:BTBAeta}) and (\ref{eq:theta0}) iteratively and to
plug back the resulting expression into (\ref{eq:E0eta}).

At leading (and subleading) order  for $L\to\infty$  the pseudo energy takes the form
\begin{equation}
\epsilon(\theta)=2mL\cosh\theta-\ln\Bigl(\frac{S(\theta)}
{S(\theta-\theta_{0})}\Bigr)\:,
\label{eq:LOepsilon}
\end{equation}
where $\theta_{0}$ is determined from the equation
\begin{equation}
1-\lambda(\theta_{0})S(\theta_{0})e^{-2mL\cosh\theta_{0}}=0\:.\label{eq:LOtheta0}
\end{equation}
 For very large $L$ the exponential term is very small and $\theta_{0}$
has to be very small as well in order to be close to the pole of the
reflection factors. Assuming $g_{+}g_{-}>0$ we find
\begin{equation}
\theta_{0}^{(0)}=\frac{i}{2}g_{+}g_{-}e^{-mL}\:.
\label{theta0as}
\end{equation}
The leading order energy comes from the non-integral term of (\ref{eq:E0eta})
as
\begin{equation}
E_{0}^{(1)}(L)=-\frac{m}{4}g_{+}g_{-}e^{-mL}\:.
\end{equation}
Now it is easy to follow what happens for the case $g_{+}g_{-}<0$.
We simply follow the movement of $\theta_{0}$ when we change the
sign of $\varphi_{-}$. This can be followed in the asymptotic solution
(\ref{theta0as}) and the result is that we have to take the choice
$\Im m(\theta_{0})<0$ for the solution of (\ref{eq:theta0}).

At next to leading order we have two sources of corrections. First,
the integral term in (\ref{eq:E0eta}) has to be expanded. Second,
$\theta_{0}$ will gain correction, too, which can be calculated by
using (\ref{eq:LOepsilon}) in (\ref{eq:LOtheta0}). We found the
correction of $\theta_{0}:$
\begin{equation}
\theta_{0}=\theta_{0}^{(0)}+\theta_{0}^{(1)}+\dots\qquad;\qquad
\theta_{0}^{(1)}=-\frac{i}{8}g_{+}^{2}g_{-}^{2}\varphi(0)e^{-2mL}\:,
\end{equation}
where we used that $\partial_{\theta}S(\theta)\vert_{0}=-i\varphi(0)$.

The energy expression at this order is:
\begin{equation}
E_{0}^{(2)}(L)=\frac{m}{8}g_{+}^{2}g_{-}^{2}\varphi(0)e^{-2mL}-
m\int_{-\infty+i\eta}^{\infty+i\eta}\frac{d\theta}{4\pi}\cosh(\theta)\,
\lambda(\theta)\, e^{-2mL\cosh(\theta)}\:,
\end{equation}
where, due to the shifted contour, the integral is convergent.

\section{Asymptotic solution}
\label{asymsol}
In this Appendix we calculate the asymptotic solution of the canonical
BTBA equations (\ref{cano}).
In the asymptotic limit the massive nodes are small,  the terms containing $L_Q$ can be neglected
and all the magnonic Y-functions are constants. Then we only need the integrals of the kernel functions:
\begin{equation}
\begin{split}
\tilde K^{mQ}_{vwx}&=\int_{-\infty}^\infty{\rm d}u\,K^{mQ}_{vwx}(u,v)=
\left\{ \begin{split} m+1\quad m&<Q\\Q\,\,\,\,\,
\quad m&\geq Q\end{split}\right. \quad ;\qquad
\tilde K_m=\int_{-\infty}^\infty{\rm d}u\,K_m(u-v)=1\\
\tilde K_{m^\prime m}&=\int_{-\infty}^\infty{\rm d}u\,K_{m^\prime m}(u-v)=
\left\{\begin{split}m^\prime&<m\quad 2m^\prime\\
m^\prime&=m\quad 2m-1\\
m^\prime&>m\quad 2m\end{split}\right.~~ ;~~~~
\tilde K_{yQ}=\int_{-2}^2{\rm d}u\,K_{yQ}(u,v)=1
\end{split}
\end{equation}
The TBA equations simplify drastically:
\begin{align}
\ln Y^o_Q&=-2\psi Q-R\tilde\varepsilon_Q+\ln M_Q+2\sum_{m=1}^{Q-1}(m+1)
{\cal L}^o_m+2Q\sum_{m=Q}^\infty{\cal L}^o_m+2{\cal L}_+,\\
\ln Y^o_+&=f-t+\sum_{m=1}^\infty\left({\cal L}^o_m-\tilde{\cal L}^o_m\right),\\
\ln Y^o_{m|v}&=2mf+\sum_{m^\prime=1}^m 2m^\prime{\cal L}^o_{m^\prime}+
2m\sum_{m^\prime=m+1}^\infty{\cal L}^o_{m^\prime}-{\cal L}^o_m,\label{Yvw}\\
\ln Y^o_{m|w}&=2mt+\sum_{m^\prime=1}^m 2m^\prime\tilde{\cal L}^o_{m^\prime}+
2m\sum_{m^\prime=m+1}^\infty\tilde{\cal L}^o_{m^\prime}-
\tilde{\cal L}^o_m.\label{Yw}
\end{align}
We can simply rewrite (\ref{Yvw}) as
\begin{equation}
\ln(1+Y^o_{m|v})=2mf+\sum_{m^\prime=1}^m 2m^\prime{\cal L}^o_{m^\prime}+
2m\sum_{m^\prime=m+1}^\infty{\cal L}^o_{m^\prime}
\end{equation}
and from this we find
\begin{equation}
\ln(1+Y^o_{m+1|v})+\ln(1+Y^o_{m-1|v})=2\ln Y^o_{m|v},
\end{equation}
and the boundary condition
\begin{equation}
Y^o_{0|v}=0.
\label{bc}
\end{equation}
The asymptotic $Y^o_{m|v}$ functions are thus solution of the constant Y-system equations
\begin{equation}
(Y^o_{m|v})^2=(1+Y^o_{m+1|v})(1+Y^o_{m-1|v})
\end{equation}
and the boundary condition (\ref{bc}). It is well known that the solution of this system is of the form
\begin{equation}
Y^o_{m|v}=\frac{\sinh pm\,\sinh p(m+2)}{\sinh^2 p},
\end{equation}
where $p$ is some parameter.
Similarly manipulating (\ref{Yw}) we find that the asymptotic $Y^o_{m|w}$
functions must be of the form
\begin{equation}
Y^o_{m|w}=\frac{\sinh \tilde pm\,\sinh \tilde p(m+2)}{\sinh^2 \tilde p},
\end{equation}

So far we have treated  infinite sums rather formally. Let us now introduce the notation
\begin{equation}
\ell_m=\frac{\sinh pm}{\sinh p}
\end{equation}
and write the cutoff sum
\begin{equation}
\sum_{m=Q}^\Lambda{\cal L}^o_m=\ell_{Q+1}-\ell_Q+\ell_{\Lambda+1}-\ell_{\Lambda+2}=
\ell_{Q+1}-\ell_Q+\ln\frac{\sinh(\Lambda+1)p}{\sinh(\Lambda+2)p}.
\end{equation}
We see that the $\Lambda\to\infty$ limit exists if  $p$ has a real part. Assuming   $p>0$ we have
\begin{equation}
\sum_{m=Q}^\infty{\cal L}^o_m=\ell_{Q+1}-\ell_Q-p.
\end{equation}
Using this formula we find that the canonical TBA equations are satisfied
if $p=f$. Completely analogous considerations lead to the conclusion that
$\tilde p=t>0$ real.

So far we have solved equations (\ref{Yvw}-\ref{Yw}). Using these results
we can calculate the asymptotic solution of the fermionic and massive
Y-functions as well. We find
\begin{equation}
Y^o_+=Y^o_-=\frac{\cosh f}{\cosh t},\qquad f>t
\end{equation}
and
\begin{equation}
Y^o_Q=4{\rm e}^{-2(f+\psi)Q}\frac{\sinh^2 fQ}{\sinh^2 f}\,(\cosh f-\cosh t)^2
\left(\frac{x^{[Q]}}{x^{[-Q]}}\right)^{2L+2}\,M_Q.
\end{equation}

\section{NLO TBA calculation}

In this Appendix we give details about the calculation of $\mathcal{M}_{Q'Q}$
from the linearized TBA equations
\begin{equation}
y_{Q}=u_{Q'}\mathcal{M}_{Q'Q}\:.
\end{equation}
We determine $\mathcal{M}_{Q'Q}$ in two different ways by generalizing
the calculations in \cite{Ahn:2011xq} for the two different deformation
angles in the $S^{5}$ and $AdS_{5}$ parts. We start with the expansion
of the TBA equations.

First we solve the recursion equation for $y_{n\vert w}$:
\begin{equation}
y_{n\vert w}=\left(\frac{[n-1]_{t}[n+1]_{t}}{[n]_{t}^{2}}y_{n-1\vert w}
+\frac{[n+1]_t[n+3]_{t}}{[n+2]_{t}^{2}}y_{n+1\vert w}\right)\star s
+\delta_{n1}c_{w}\star s\,,
\end{equation}
 where
\begin{equation}
c_{w}=\frac{y_{+}-y_{-}}{1-Y^o_{+}}\left(\Theta(u+2)-\Theta(u-2)\right)
=\frac{[2]_{t}}{[2]_{t}-[2]_{f}}2\pi u_{Q}K_{Qy}\left(\Theta(u+2)-
\Theta(u-2)\right)\label{c1}
\end{equation}
and $\Theta$ is the unitstep function. We use Fourier transformation,
where $\tilde{s}=(2\cosh\frac{\omega}{g})^{-1}=(k+k^{-1})^{-1}$ with
$k\equiv e^{-\frac{|\omega|}{g}}$. The solution which decreases for
large $n$ (to respect the asymptotics of  $Y_{n\vert w}$) and is compatible
with the $\delta_{n,1}$ term is
\begin{equation}
\tilde{y}_{n\vert w}=\frac{\tilde{c}_{w}k}{[2]_{t}}\left(\frac{[n+1]_{t}}
{[n]_{t}}k^{n-1}-\frac{[n+1]_{t}}{[n+2]_{t}}k^{n+1}\right)\:.
\end{equation}
Then we solve the recursion for $y_{n\vert v}$:
\begin{equation}
y_{n\vert v}=\frac{[n-1]_{f}[n+1]_{f}}{[n]_{f}^{2}}y_{n-1\vert v}\star s
+\frac{[n+1]_{f}[n+3]_{f}}{[n+2]_{f}^{2}}y_{n+1\vert v}\star s-2\pi u_{n+1}
\star s+\delta_{n1}c_{v}\star s
\end{equation}
\begin{equation}
c_{v}=\frac{y_{-}-y_{+}}{1-\frac{1}{Y^o_{+}}}\left(\Theta(u+2)
-\Theta(u-2)\right)=\frac{[2]_{f}}{[2]_{t}-[2]_{f}}2\pi u_{Q}K_{Qy}
\left(\Theta(u+2)-\Theta(u-2)\right)\:.
\end{equation}
In Fourier space it takes the form
\begin{equation}
(k+k^{-1})\tilde{y}_{n\vert v}=\frac{[n-1]_{f}[n+1]_{f}}{[n]_{f}^{2}}
\tilde{y}_{n-1\vert v}+\frac{[n+1]_{f}[n+3]_{f}}{[n+2]_{f}^{2}}
\tilde{y}_{n+1\vert v}-\tilde{\mathcal{S}}_{n+1}+\delta_{n1}\tilde{c}_{v}
\end{equation}
with some inhomogeneous source terms $\mathcal{S}$. The solution of
the inhomogeneous equation is provided by carefully choosing the combination
of the solutions of the homogenous equation
\begin{eqnarray}
\tilde{y}_{N\vert v} & = & \Bigl(\frac{[n+1]_f}{[n]_f}k^{n-1}
-\frac{[n+1]_f}{[n+2]_f}k^{n+1}\Bigr)\Bigl(A_{-}-c
\sum_{j=1}^{n}\frac{\tilde{\mathcal{S}}_{j+1}k^{-j-2}(k^{-2}[j]_{f}-[j+2]_{f})}
{[j+1]_{f}}\Bigr)\\
 & +& \Bigl(\frac{[n+1]_f}{[n]_f}k^{1-n}-\frac{[n+1]_f}{[n+2]_f}
 k^{-n-1}
 \Bigr)\Bigl(A_{+}-c\sum_{j=1}^n\frac{\tilde{\mathcal{S}}_{j+1}k^{j-2}
 (k^{-2}[j+2]_{f}-[j]_{f})}{[j+1]_{f}}\Bigr)
\end{eqnarray}
where
\[
c^{-1}=(k^{-2}-1)(qk^{-2}-q^{-1})(q^{-1}k^{-2}-q)\qquad;\qquad q=e^{f}
\]
and $A_{\pm}$ should be fixed from the boundary conditions. In order
to have the decreasing asymptotics at $n\to\infty$ we need to take
\begin{equation}
A_{+}=c\sum_{j=1}^{\infty}\frac{\tilde{\mathcal{S}}_{j+1}k^{j-2}
\left(k^{-2}[j+2]_{f}-[j]_{f}\right)}{[j+1]_{f}}\:.
\end{equation}
and from the starting $n=1$ value we found
\begin{equation}
A_{-}=k\left(\frac{\tilde{c}_{v}}{[2]_{f}}-A_{+}k\right)\:.
\end{equation}
Once $y_{m\vert v}$ and $y_{n\vert w}$ are known we can plug back
their expression into
\begin{eqnarray}
y_{+}+y_{-} & = & 2\left(A_{1\vert v}y_{1|v}-A_{1\vert w}y_{1|w}\right)
\star s
+4\pi u_{Q}K_{xv}^{Q1}\star s-2\pi u_{Q}K_{Q}\\
y_{+}-y_{-} & = & 2\pi u_{Q}K_{Qy}\:. \nonumber
\end{eqnarray}
With the help of these magnonic nodes the full $u_{Q}$ contribution
to the NLO L\"uscher correction turns out to be
\begin{eqnarray}
\frac{1}{2\pi}y_{Q} & = & u_{Q'}K_{sl(2)}^{Q'Q}+u_{Q'}2s\star
 K_{vx}^{Q'-1,Q}
+2\bigl[A_{1\vert v}y_{1|v}\star s\hat{\star}K_{yQ}
+A_{Q-1\vert v}
y_{Q-1|v}\star s\,
\,\,\,\,\,\,\,\\
 &  & \,\,\,\,\,\,\,\,\,\,-\frac{u_{Q}K_{Qy}}{2(Y^o_{+}-1)}\hat{\star}(K_{Q}
 -s\star K_{yQ})+\frac{u_{Q}K_{Qy}}{1-\frac{1}{Y^o_{+}}}\hat{\star}s
 \star K_{vx}^{1Q}+\frac{y_{-}}{(Y^o_{+}-1)}\hat{\star}s\star K_{yQ}\bigr]\:.
 \nonumber
\end{eqnarray}
We plug back the solution for $y_{-}$ in terms of $y_{1\vert v}$
and $y_{1\vert w}$, which  can be further reexpressed in terms of $u_{Q}$
. After similar simplifications to \cite{Ahn:2011xq} we obtain the
solution in a relatively simple form
\begin{eqnarray}
y_{Q_{2}} & = & u_{Q_{1}}\Bigg\{2\pi K_{sl(2)}^{Q_{1}Q_{2}}+4\pi
\sum_{j=0}^{Q_{1}-2}K_{Q_{2}-Q_{1}+2j+1}\star s+\frac{2[2]_{t}}{[2]_{f}-[2]_{t}}
\frac{1}{i}\partial_{u_{1}}\ln a_{1}^{Q_{1}Q_{2}}(u_{1},u_{2}) \nonumber \\
 & + & \frac{2}{[2]_{f}-[2]_{t}}\frac{1}{i}\partial_{u_{1}}
 \left[\frac{[Q_{2}-1]_{f}}{[Q_{2}]_{f}}\ln a_{2}^{Q_{1}Q_{2}}(u_{1},u_{2})
 +\frac{[Q_{1}-1]_{f}}{[Q_{1}]_{f}}\ln a_{2}^{Q_{2}Q_{1}}(u_{2},u_{1})^{\star}
 \right] \\
 & + & \frac{4\pi}{[Q_{1}]_{f}[Q_{2}]_{f}}\sum_{k=0}^{Q_{1}-1}[k]_{f}[k-Q_{1}]_{f}
 \left[[Q_{2}+1]_{f}K_{Q_{2}-Q_{1}+2k-1}-[Q_{2}-1]_{f}K_{Q_{2}-Q_{1}+2k+1}
 \right]\star s\Bigg\}\,, \nonumber~~~~~~\label{yQ1}
\end{eqnarray}
where we introduced the functions (\ref{eq:a1a2}).

There is an alternative calculation for the same matrix, $\mathcal{M}$,
based on the scattering description of the double wrapping correction
of the ground-state energy for the theory in which both $su(2)$ factors
are deformed by the twist matrix
\begin{equation}
\Gamma=e^{f J+tR}\otimes e^{f J+t R}
\end{equation}
In calculating the NLO Luscher correction of the ground state energy,
following \cite{Ahn:2011xq}, we have to evaluate
\begin{eqnarray}
\mathcal{M} & = & -\frac{i\partial_{1}\mbox{sTr}_{12}(\Gamma_{12}\ln S_{12})}
{([2]_{f}-[2]_{t})^{4}[Q_{1}]_{f}^{2}[Q_{2}]_{f}^{2}}\nonumber \\
 & = & 2\pi K_{sl(2)}-2\frac{i\partial_{1}\mbox{sTr}(\Gamma_{Q_{1}Q_{2}}
 \ln S^{Q_{1}Q_{2}})}{([2]_{f}-[2]_{t})^{2}[Q_{1}]_{f}[Q_{2}]_{f}}
\end{eqnarray}
where we used the factorization
\begin{equation}
\ln S_{12}=(\ln S_{0})\mathbb{I}\otimes\mathbb{I}+\ln S_{su(2\vert2)}
\otimes\mathbb{I}+\mathbb{I}\otimes\ln S_{su(2\vert2)}
\end{equation}
and explicitly evaluated the supertrace
\begin{equation}
\mbox{sTr}_{12}(e^{f J+tR})=([2]_f-[2]_{t})^{2}[Q_{1}]_{f}[Q_{2}]_{f}\:.
\end{equation}
We focus on the $Q_{2}\geq Q_{1}$ subspace for one $su(2\vert2)$
factor. Decomposing the trace with respect to $su(2)\otimes su(2)\subset su(2\vert2)$
we can write
\begin{eqnarray}
([2]_{f}-[2]_{t})^{2}[Q_{1}]_{f}[Q_{2}]_{f}
\mbox{\ensuremath{\mathcal{M}_{Q_{1}Q_{2}}}} & =
& -i\partial_{1}\mbox{sTr}(\Gamma_{Q_{1}Q_{2}}\ln S^{Q_{1}Q_{2}})\\
 & = & -i\partial_{1}\sum_{s_{L},s_{R}}\mbox{sTr}(e^{f J}\otimes
 e^{t R}\ln S^{Q_{1}Q_{2}}(s_{L},s_{R}))\nonumber \\
 & = & -i\partial_{1}\sum_{s_{L},s_{R}}(-1)^{2s_{R}}[2s_{R}+1]_{t}[2s_{L}+1]_{f}
 \ln\det S^{Q_{1}Q_{2}}(s_{L},s_{R})\:.\nonumber
\end{eqnarray}
where we sum all possible left and right spins $s_{L},s_{R}$ for
a given $Q_{1},Q_{2}$. The twist factors are
\begin{equation}
\mbox{Tr}_{s_{L}}(e^{f J})=(-1)^{2s_{L}}[2s_{L}+1]_{f}\qquad;\qquad\mbox{sTr}_{s_{R}}
(e^{t R})
=(-1)^{2s_{R}}[2s_{R}+1]_{t}
\end{equation}
Fortunately the $\ln\det S^{Q_{1}Q_{2}}(s_{L}s_{R})$ pieces were
calculated in \cite{Ahn:2011xq} and now we just put them together
to calculate the combination
\begin{eqnarray}
&& [Q_{1}]_{f}[Q_{2}]_{f}\ensuremath{(\mathcal{M}_{Q_{1}Q_{2}}
-2\pi K_{sl_{2}}^{Q_{1}Q_{2}})} = -\mathcal{K}_{f}^{Q_{1}Q_{2}}
 +([2]_{t}-[2]_{f})^{-1}
 [Q_{21}]_{f}\frac{1}{i}\partial_{1}\ln(B)\\
 &  & \hspace{1cm} +([2]_{t}-[2]_{f})^{-1}\left([Q_{1}]_{f}[Q_{2}-1]_{f}+[Q_{1}-1]_{f}[Q_{2}]_{f}
 -[2]_{t}[Q_{1}]_{f}[Q_{2}]_{f}\right)\frac{1}{i}\partial_{1}\ln(A)\nonumber
\end{eqnarray}
where $Q_{21}=Q_{2}-Q_{1}$, $\hat{Q}_{21}=Q_{2}+Q_{1}$ and we used
the quantities (\ref{eq:AB}) and (\ref{eq:Kf}). This expression
gives the same matrix $\mathcal{M}$ as (\ref{yQ1}).

\section{Details of the weak coupling expansion}

In performing the weak coupling expansion of the various terms we
use the conventions of \cite{Ahn:2011xq}:
\begin{equation}
x^{[\pm Q]}=\frac{\tilde{P}-iQ}{2g}\left(\sqrt{1+\frac{4g^{2}}{\tilde{P}^{2}+Q^{2}}}\mp1\right)\qquad;
\qquad u\pm\frac{iQ}{g}=x^{[\pm Q]}+\frac{1}{x^{[\pm Q]}}
\end{equation}
In the expansion of $u_{Q}^{(0)}$ we need to keep the second order
terms

\begin{equation}
\frac{\mathrm{d}\tilde{P}_{Q}}{\mathrm{d}u}(0)=\frac{gQ}{\sqrt{Q^{2}+4g^{2}}}=
g\left(1-2\frac{g^{2}}{Q^{2}}\right)+\dots
\end{equation}
\begin{equation}
e^{-\tilde{\epsilon}_{Q}(0)}=\frac{\sqrt{4g^{2}+Q^{2}}-Q}{\sqrt{4g^{2}+Q^{2}}+Q}=
\frac{g^{2}}{Q^{2}}\left(1-2\frac{g^{2}}{Q^{2}}\right)+\dots
\end{equation}
\begin{equation}
e^{2i(\Phi(x^{[-Q]})+\Phi(1/x^{[Q]})-\Phi(0))}=1+\frac{2\pi^{2}g^{2}}{3}+\dots
\end{equation}
\begin{equation}
m_{Q}=\sqrt{\lim_{u\to0}u^{2}M_{Q}(u)}=\sqrt{\lim_{u\to0}u^{2}\frac{\pi^{2}(g^{2}u^{2}+Q^{2})}
{\sinh^{2}\pi gu}}e^{-i\Phi(0)}+\dots=\frac{Q}{g}+\frac{\pi^{2}Qg}{3}+\dots
\end{equation}
This results in

\begin{equation}
u_{Q}^{(0)}=\frac{g}{Q}[Q]_{f}([2]_{f}-[2]_{t})e^{-(f+\Psi)Q}\left(1+g^{2}\left(\frac{\pi^{2}}{3}
-\frac{2}{Q^{2}}\right)+O(g^{4})\right)
\end{equation}
In the integral term we need to expand at leading order the various
terms:
\begin{equation}
M_{Q}(u)=\frac{\pi^{2}(g^{2}u^{2}+Q^{2})}{\sinh^{2}\pi gu}+\dots
\end{equation}
\begin{equation}
e^{-\tilde{\epsilon}_{Q}(\tilde{P})}=\frac{x^{[+Q]}}{x^{[-Q]}}=\frac{g^{2}}{\tilde{P}^{2}+Q^{2}}+\dots
=\frac{g^{2}}{g^{2}u^{2}+Q^{2}}+\dots
\end{equation}
In the term $\mathcal{M}$ we use
\begin{eqnarray}
-i\partial_{u_{1}}\log(A)|_{u=0} & = & \frac{g}{Q_{1}}+\dots\\
-i\partial_{u_{1}}\log(B)|_{u=0} & = & -\frac{g}{Q_{1}}+\dots\\
\mathcal{K}_{f}^{Q_{1}Q_{2}}\vert_{u=0} & = & 2\sum_{j=1}^{Q_{1}-1}[j]_{f}[Q_{21}+j]_{f}\frac{1}{Q_{21}+2j}\:.
\end{eqnarray}

The weak coupling expansion of the dressing phase is
\begin{equation}
K_{sl_{2}}^{Q_{1}Q_{2}}=\frac{1}{2\pi i}\partial_{\tilde{P}_{1}}\log S_{sl(2)}^{Q_{1}Q_{2}}
(\tilde{P}_{1},\tilde{P}_{2})=-K_{Q_{1}Q_{2}}-\frac{1}{\pi i}\partial_{\tilde{P}_{1}}\log
\Sigma^{Q_{1}Q_{2}}(\tilde{P}_{1},\tilde{P}_{2})\,,
\end{equation}
\begin{equation}
\frac{1}{\pi i}\partial_{\tilde{P}_{1}}\log\Sigma^{Q_{1}Q_{2}}(\tilde{P}_{1},\tilde{P}_{2})=
\frac{1}{2\pi}\biggr[\psi\left(1-\tfrac{i}{2}(\tilde{P}_{1}+iQ_{1})\right)-\psi\left(1+\tfrac{i}{2}(\tilde{P}_{21}
-i(Q_{1}+Q_{2}))\right)+c.c\biggl]
\end{equation}
where $\tilde P_{21}=\tilde P_2 -\tilde P_1$ and
$\psi(x)=\partial_x\log\Gamma(x)$
is the polygamma function. The $su(2)$ scalar factor results in
\begin{eqnarray}
-K_{su(2)}^{Q_{1}Q_{2}}=-K_{Q_{1}Q_{2}} & = & \frac{1}{4\pi}\biggr[\psi\left(\tfrac{i}{2}(\tilde{P}_{21}
-i(Q_{1}-Q_{2}))\right)+\psi\left(1+\tfrac{i}{2}(\tilde{P}_{21}-i(Q_{1}-Q_{2}))\right)\\
 &  & \qquad-\psi\left(\tfrac{i}{2}(\tilde{P}_{21}-i(Q_{1}+Q_{2}))\right)-\psi\left(1+\tfrac{i}{2}
 (\tilde{P}_{21}-i(Q_{1}+Q_{2}))\right)+c.c\biggl]\nonumber
\end{eqnarray}
 and we will need these expressions at $\tilde{P}_{1}=\tilde{P}_{2}=0$:
\begin{equation}
2\pi K_{sl_{2}}^{Q_{1}Q_{2}}=\frac{2}{Q_{2}-Q_{1}}+\frac{2}{Q_{2}+Q_{1}}+2\psi\left(\tfrac{1}{2}(Q_{2}-Q_{1})\right)
-2\psi\left(1+\tfrac{1}{2}Q_{1}\right)\,.
\end{equation}
This expression is valid for $Q_{2}>Q_{1}$. For $Q_{2}=Q_{1}$ special
care is needed and we found
\begin{equation}
2\pi K_{sl_{2}}^{QQ}=\frac{1}{Q}+2\psi(1)-2\psi\left(1+\tfrac{1}{2}Q_{1}\right)\,.
\end{equation}
Using these leading order weak coupling formulas we calculate

\begin{equation}
M^{(2)}:=-\frac{g}{4}\sum_{Q_{1},Q_{2}=1}^{\infty}u_{Q_{1}}\Bigg\{2\pi K_{sl(2)}^{Q_{1}Q_{2}}+
\frac{2}{[Q_{1}]_{f}[Q_{2}]_{f}}\frac{1}{i}{\cal K}_{f}^{Q_{1}Q_{2}}-\frac{2}{i}\partial_{u_{1}}
\log a_{1}^{Q_{1}Q_{2}}(u_{1},u_{2})\biggr\} u_{Q_{2}}
\end{equation}
After some cancellation we found it useful to group the remaining
terms in the following way:
\begin{equation}
M^{(2)}=-g^{4}\frac{(\cosh f-\cosh t)^{2}}{(\sinh f)^{2}}\tilde{\gamma}^{(2)}\,,\quad\tilde{\gamma}^{(2)}
=(A+B^{+}+B^{-}+C+D+E+F+G+X)\,,
\end{equation}
\begin{eqnarray*}
A & = & 4\sum_{Q_{1}<Q_{2}}^{\infty}\frac{1}{Q_{1}Q_{2}}\sum_{j=1}^{Q_{1}-1}
\frac{\cosh(f(Q_{2}-Q_{1}+2j))}{(Q_{2}-Q_{1}+2j)}\,,\\
B^{\pm} & = & 2\sum_{Q_{1}<Q_{2}}^{\infty}\frac{1}{Q_{1}Q_{2}}\frac{\cosh(f(Q_{2}\pm Q_{1}))}{Q_{2}\pm Q_{1}}\,,\\
C & = & \sum_{Q=1}^{\infty}\frac{1}{Q^{2}}\sum_{j=1}^{Q-1}\frac{\cosh(2jf)-1}{j}\,,\\
D & = & \frac{1}{2}\sum_{Q=1}^{\infty}\frac{\cosh(2Qf)-1}{Q^{3}}\,,\\
E & = & \psi(1)\sum_{Q=1}^{\infty}\frac{\cosh(2Qf)-1}{Q^{2}}\,,\\
F & = & -\sum_{Q_{2}=1}^{\infty}\frac{\sinh Q_{2}f}{Q_{2}}\sum_{Q_{1}=1}^{\infty}2
\frac{\sinh Q_{1}f}{Q_{1}}\psi(1+\frac{Q_1}{2})\,,\\
G & = & \sum_{Q_{2}=1}^{\infty}\frac{\sinh Q_{2}f}{Q_{2}}\sum_{Q_{1}=1}^{\infty}2
\frac{\sinh Q_{1}f}{Q_{1}^{2}}\,,\\
X & = & \sum_{Q_{1}<Q_{2}}^{\infty}\frac{1}{Q_{1}Q_{2}}\left\{ \cosh(f(Q_{2}+Q_{1}))h(Q_{2}-Q_{1})
-\cosh(f(Q_{2}-Q_{1}))h(Q_{2}+Q_{1})\right\}\,,
\end{eqnarray*}
where
\[
h(x)=\psi(\frac{x}{2})+\psi(1+\frac{x}{2})=2\psi(1+\frac{x}{2})-\frac{2}{x}\,.
\]
We found that with $f=\pm i(\pi-\phi)$
\[
(A+B^{+}+B^{-}+C+D)^{\prime}=-\frac{\phi^{2}}{2}\cot\phi\,,
\]
where the derivative is with respect to $\phi$: $f^{\prime}(\phi)=\frac{df(\phi)}{d\phi}$.
\[
E^{\prime}=2\psi(1)(\phi-\frac{\pi}{2})\,,
\]
\begin{equation}
\begin{aligned}
G =-\phi S_{2}(\phi)\,,\qquad
 & S_{2}(\phi) =\sum_{Q=1}^{\infty}\frac{\sin(\pi-\phi)Q}{Q^{2}}
=\int_{0}^{\phi}{\rm d}y\log(2\cos\frac{y}{2})\,,
\\
F=\frac{\phi}{2}\tilde{S}_{1}(\phi)\,,\qquad & \tilde{S}_{1}(\phi)  =2
\sum_{Q=1}^{\infty}\frac{\sin(\pi-\phi)Q}{Q}\psi(1+\frac{Q}{2})\:.
\end{aligned}
\end{equation}

In calculating $X$ we change summation from $Q_{1},Q_{2}$ to $Q_{2}+Q_{1}=m$
and $Q_{2}-Q_{1}=n$ keeping in mind that $m$ and $n$ must have
the same parity:
\[
X=\sum_{n=1}^{\infty}\sum_{m>n}^{\infty}\frac{4}{m^{2}-n^{2}}(\cosh(fm)h(n)-\cosh(fn)h(m))
=\sum_{n=1}^{\infty}\sum_{m\neq n}^{\infty}\frac{4}{m^{2}-n^{2}}\cosh(fm)h(n)\:.
\]
 By changing to $f=\pm i(\pi-\phi)$, separating the even and odd
contributions and using the formulas
\begin{equation}
\sum_{m:m\neq n}^{\infty}\frac{\cos mx}{m^{2}-n^{2}}=\begin{cases}
\frac{1}{2n^{2}}+\frac{\cos nx}{4n^{2}}+\frac{(x-\pi)\sin nx}{2n} & \mbox{\qquad if \ensuremath{n\in\mathbb{Z}}}\\
\frac{1}{2n^{2}}-\frac{\pi}{2n}\frac{\cos n(\pi-x)}{\sin\pi n} & \qquad\mbox{if }\ensuremath{n\notin\mathbb{Z}}
\end{cases}
\end{equation}
we found that
\begin{equation}
X^{\prime}=(\pi-2\phi)\tilde{h}^{\prime}-\tilde{h}\,, \qquad\tilde{h}=\sum h(n)\frac{\sin n(\pi-\phi)}{n}
=\tilde{S}_{1}(\phi)-2S_{2}(\phi)\:.
\end{equation}
Clearly
\begin{equation}
\tilde{h}(0)=0\,,\qquad\tilde{h}^{\prime}(\phi)=\psi(1)-\phi\cot\phi\:.
\end{equation}
Collecting the terms together
\begin{equation}
\tilde{\gamma}^{(2)\prime}=-\frac{\phi^{2}}{2}\cot\phi+2\psi(1)(\phi-\frac{\pi}{2})
-\frac{\tilde{h}}{2}+(\pi-\frac{3\phi}{2})\tilde{h}^{\prime}\:.
\end{equation}
In calculating $\tilde{S}_{1}(\phi)$ we calculate its derivative
using that
\begin{equation}
\sum_{m=1}^{\infty}z^{m-1}\psi(1+\frac{m}{2})=\frac{2\log(1-z)}{z(z^{2}-1)}+\frac{\psi(1)}{1-z}
+\frac{2\log2}{z^{2}-1}\:.
\end{equation}
So
\begin{equation}
\tilde{S}_{1}(\phi)^{\prime}=2\log(2\cos\frac{\phi}{2})-\phi\cot\phi+\psi(1)\:.
\end{equation}
This implies that combining $\tilde{\gamma}^{(2)\prime}$ with the
derivative of the integral term we have
\begin{equation}
\gamma_{2}^{(2)\prime}(0)=0\,,\qquad\gamma_{2}^{(2)\prime\prime}(\phi)=\frac{1}{2}\psi(1)
-\frac{1}{2}\tilde{h}^{\prime}=\frac{\phi}{2}\cot\phi\:,
\end{equation}
which agrees with the gauge theory result.

\providecommand{\href}[2]{#2}\begingroup\raggedright\endgroup

\end{document}